\documentclass[twocolumn]{aastex631}

\usepackage{array}
\usepackage{amsmath}
\usepackage{xcolor}
\usepackage{multirow}
\usepackage{booktabs}
\usepackage{soul}
\usepackage{pifont}
\usepackage{makecell}
\definecolor{ao}{rgb}{0.0, 0.5, 0.0}
\definecolor{carnelian}{rgb}{0.7, 0.11, 0.11}
\definecolor{mikadoyellow}{rgb}{1.0, 0.77, 0.05}

\shorttitle{First Estimate of the Ambipolar Diffusivity Coefficient in HOPS-370}
\shortauthors{Thieme et al.}
\graphicspath{{./}{figures/}}

\begin{document}

\title{The First Estimation of the Ambipolar Diffusivity Coefficient from Multi-Scale Observations of the Class 0/I Protostar, HOPS-370}

\correspondingauthor{Travis J. Thieme}
\email{tjthieme@asiaa.sinica.edu.tw}

\author[0000-0003-0334-1583]{Travis J. Thieme}
\affiliation{Institute of Astronomy, National Tsing Hua University, No. 101, Section 2, Kuang-Fu Rd., Hsinchu 30013, Taiwan}
\affiliation{Center for Informatics and Computation in Astronomy, National Tsing Hua University, No. 101, Section 2, Kuang-Fu Rd., Hsinchu 30013, Taiwan}
\affiliation{Department of Physics, National Tsing Hua University, No. 101, Section 2, Kuang-Fu Rd., Hsinchu 30013, Taiwan}
\affiliation{Academia Sinica Institute of Astronomy and Astrophysics,
11F of Astronomy-Mathematics Building, AS/NTU, No.1, Sec. 4, Roosevelt Rd, Taipei 10617, Taiwan}

\author[0000-0001-5522-486X]{Shih-Ping Lai}
\affiliation{Institute of Astronomy, National Tsing Hua University, No. 101, Section 2, Kuang-Fu Rd., Hsinchu 30013, Taiwan}
\affiliation{Center for Informatics and Computation in Astronomy, National Tsing Hua University, No. 101, Section 2, Kuang-Fu Rd., Hsinchu 30013, Taiwan}
\affiliation{Department of Physics, National Tsing Hua University, No. 101, Section 2, Kuang-Fu Rd., Hsinchu 30013, Taiwan}
\affiliation{Academia Sinica Institute of Astronomy and Astrophysics,
11F of Astronomy-Mathematics Building, AS/NTU, No.1, Sec. 4, Roosevelt Rd, Taipei 10617, Taiwan}

\author[0000-0003-3497-2329]{Yueh-Ning Lee}
\affiliation{Department of Earth Sciences, National Taiwan Normal University, Taipei 116059, Taiwan}
\affiliation{Center of Astronomy and Gravitation, National Taiwan Normal University, Taipei 116059, Taiwan}
\affiliation{Physics Division, National Center for Theoretical Sciences, Taipei 106319, Taiwan}

\author[0000-0002-6868-4483]{Sheng-Jun Lin}
\affiliation{Academia Sinica Institute of Astronomy and Astrophysics, 11F of Astronomy-Mathematics Building, AS/NTU, No.1, Sec. 4, Roosevelt Rd, Taipei 10617, Taiwan}

\author[0000-0003-1412-893X]{Hsi-Wei Yen}
\affiliation{Academia Sinica Institute of Astronomy and Astrophysics, 11F of Astronomy-Mathematics Building, AS/NTU, No.1, Sec. 4, Roosevelt Rd, Taipei 10617, Taiwan}



\begin{abstract}

Protostars are born in magnetized environments. 
As a consequence, the formation of protostellar disks can be suppressed by the magnetic field efficiently removing angular momentum of the infalling material.
Non-ideal MHD effects are proposed to as one way to allow protostellar disks to form. 
Thus, it is important to understand their contributions in observations of protostellar systems. 
We derive an analytical equation to estimate the ambipolar diffusivity coefficient at the edge of the protostellar disk in the Class 0/I protostar, HOPS-370, for the first time, under the assumption that the disk radius is set by ambipolar diffusion.
Using previous results of the protostellar mass, disk mass, disk radius, density and temperature profiles and magnetic field strength, we estimate the ambipolar diffusivity coefficient to be $1.7^{+1.5}_{-1.4}\times10^{19}\,\mathrm{cm^{2}\,s^{-1}}$. 
We quantify the contribution of ambipolar diffusion by estimating its dimensionless Els\"{a}sser number to be $\sim1.7^{+1.0}_{-1.0}$, indicating its dynamical importance in this region.
We compare to chemical calculations of the ambipolar diffusivity coefficient using the Non-Ideal magnetohydrodynamics Coefficients and Ionisation Library (NICIL), which is consistent with our results.
In addition, we compare our derived ambipolar diffusivity coefficient to the diffusivity coefficients for Ohmic dissipation and the Hall effect, and find ambipolar diffusion is dominant in our density regime.
These results demonstrate a new methodology to understand non-ideal MHD effects in observations of protostellar disks. 
More detailed modeling of the magnetic field, envelope and microphysics, along with a larger sample of protostellar systems is needed to further understand the contributions of non-ideal MHD. 

\end{abstract}



\section{Introduction} \label{sec:introduction}

It has long been known that magnetic fields  (B-fields) play a critical role in regulating the formation of protostellar disks around low-mass protostars \citep[e.g.,][]{Mestel1956MNRAS.116..503M}.
Molecular cloud cores are observed to be strongly magnetized, with normalized mass-to-flux ratios of $\mu\sim2-10$ \citep[e.g.,][]{Crutcher1999ApJ...520..706C, Troland2008ApJ...680..457T, Crutcher2012ARA&A..50...29C}.
Early ideal magnetohydrodynamic (MHD) simulations show that rotationally supported disks (RSDs) could not form due to magnetic braking efficiently transferring angular momentum away from the collapsing central region in magnetized ($\mu \leq 10$) dense cores \citep[e.g.,][]{Allen2003ApJ...599..351A,Matsumoto2004ApJ...616..266M,Banerjee2006ApJ...641..949B,Price2007Ap&SS.311...75P,Hennebelle2008A&A...477....9H,Mellon2008ApJ...681.1356M, Joos2012A&A...543A.128J}.
However, observational studies revealed the presence of rotationally-supported Keplerian disks around several young, highly-embedded protostars \citep[e.g.,][]{Tobin2012Natur.492...83T, Murillo2013A&A...560A.103M, Lee2014ApJ...786..114L,Yen2017ApJ...834..178Y,Ohashi2023ApJ...951....8O}. 
This contradiction between observations and simulations was coined the so-called \textit{``Magnetic Braking Catastrophe''} and raised the fundamental question of \textit{how could these protostellar disks form in such magnetized environments}?

Non-ideal MHD effects, namely ambipolar diffusion (AD), Ohmic dissipation (OD) and the Hall effect (HE), have been suggested as one possible route to overcome magnetic braking and form a rotationally-supported protostellar disk \citep[e.g.,][]{Inutsuka2010ApJ...718L..58I,Li2011ApJ...738..180L,Braiding2012MNRAS.427.3188B,Kengo2015ApJ...801..117T,Wurster2016MNRAS.457.1037W,Wurster2019MNRAS.489.1719W,Wurster2020MNRAS.495.3795W,Wurster2021MNRAS.507.2354W}. 
These non-ideal MHD terms describe the various regimes of coupling of the ions, electrons and charged grains to the magnetic field, as well as their interactions with the neutral particles \citep[e.g.,][see the recent reviews by \citealt{Wurster2018FrASS...5...39W, Zhao2020SSRv..216...43Z,Tsukamoto2023ASPC..534..317T}]{Wardle1999MNRAS.303..239W,Nakano2002ApJ...573..199N}.
In terms of relative importance, Ohmic dissipation is efficient at high densities, such as the midplane of a protostellar disk, while the Hall effect and ambipolar diffusion are more efficient at  intermediate and low densities, respectively, such as the upper disk layers and in the protostellar envelope \citep[e.g.,][]{Marchand2016A&A...592A..18M,Wurster2018MNRAS.475.1859W,Wurster2021MNRAS.501.5873W}.
However, the Hall effect seems to be transient does not last for long after the formation of a protostellar disk \citep{Zhao2020SSRv..216...43Z,Lee2021ApJ...922...36L}.
While simulations clearly show the importance of non-ideal MHD effects in the formation and evolution of protostellar disks, they have yet to be quantified observationally. 
\citet{Yen2018A&A...615A..58Y} attempted to observe the velocity drift between ions and neutral particles (ambipolar diffusion) in the infalling envelope of a young Class 0 protostar, B335. 
However, no velocity drift was detected and thus, it is important to look into other possibilities on how non-ideal MHD effects can be quantified observationally. 

In this paper, we aim to understand the role of ambipolar diffusion in protostellar disk formation by using a methodology first developed by \citet{Hennebelle2016ApJ...830L...8H}, and later revisited by \citet{Lee2021ApJ...922...36L,Lee2024ApJ...961L..28L}.
This methodology leads to an analytical equation describing the expected protostellar properties, in particular the protostellar disk radius, due to ambipolar diffusion \citep{Hennebelle2016ApJ...830L...8H}. 
The disk radius estimated with this analytical equation ($R_\mathrm{AD}$) was found to be in good agreement with the disk radius estimated from MHD simulations ($R_{\mathrm{sim}}$), with $R_{\mathrm{sim}}/R_\mathrm{AD}\sim1$ \citep{Hennebelle2016ApJ...830L...8H,Hennebelle2020A&A...635A..67H,Commercon2022A&A...658A..52C}. 
Thus, by backwards engineering the equation, we can estimate the ambipolar diffusivity coefficient, $\eta_\mathrm{AD}$, from observable quantities under certain assumptions.
Using multi-scale observations of the young protostar, HOPS-370, we present a methodology to estimate the ambipolar diffusivity coefficient for the first time, in order to understand the role of ambipolar diffusion in the formation and evolution of protostellar disks. 

HOPS-370 is a Class 0/I protostar in the Orion A molecular cloud (D$=392.8\,\mathrm{pc}$; \citealt{Tobin2020ApJ...890..130T}). 
Observations from the Herschel Orion Protostar Survey (HOPS) constrain the bolometric luminosity ($L_\mathrm{bol}$) and temperature ($T_\mathrm{bol}$) to be 314$\,L_\odot$ and 71.5$\,$K, respectively \citep{Furlan2016ApJS..224....5F}. 
The protostellar mass and disk properties were extensively studied by \citet{Tobin2020ApJ...905..162T} as part of the VLA/ALMA Nascent Disk and Multiplicity (VANDAM) Survey of Orion Protostars.
By using MCMC radiative transfer modeling to fit the dust continuum and several molecular lines, they found an average disk radius of $94\,\mathrm{au}$, an average protostellar mass of $2.5\,M_\odot$, and a disk mass of $0.035\,M_\odot$. 
More recently, Kao \& Yen et al. (in prep.) have derived the core-scale plane-of-sky magnetic field strength to be $B_\mathrm{pos}=0.51\,\mathrm{mG}$.
The combination of these derived properties make HOPS-370 an ideal candidate for an initial study on the role ambipolar diffusion plays in this source using this new methodology.

This paper is organized as follows. 
In Section \ref{sec:methods}, we describe our methodology and assumptions to estimate $\eta_\mathrm{AD}$ at the edge of the HOPS-370 protostellar disk.
Our resulting value of $\eta_\mathrm{AD}$ and a comparison with a more theoretical non-ideal MHD estimate is given in Section \ref{sec:results}.
Several implications and uncertainties are discussed in Section \ref{sec:discussion}. 
Section \ref{sec:conclusion} summarizes our main results and discussions.

\section{Methods}\label{sec:methods}
\subsection{The Relation Between Protostellar Disk Properties and the Ambipolar Diffusivity Coefficient}\label{sec:relation}

Here, we present an analytical equation that relates properties of the protostellar disk at the disk-envelope interface to the ambipolar diffusion coefficient. 
\citet{Hennebelle2016ApJ...830L...8H} were the first to derive such an equation, however they make a number of simplifications to remove terms related to the density and temperature, which differs from the modeling of HOPS-370.
A derivation is provided in Appendix \ref{sec:etaderivation}, while a summary and overview is presented here.
The main assumptions in this derivation are that
\begin{enumerate}
\item ambipolar diffusion is the main diffusion process,
\item the angular momentum is counteracted by magnetic braking resulting in the advection and braking timescales to be of the same order,
\item the toroidal field generated by differential rotation is offset by the ambipolar diffusion in the vertical direction resulting in the Faraday induction and vertical diffusion timescales to be of the same order, 
\item infalling and rotational velocities of the gas near the disk edge both scale with the Keplerian velocity, and 
\item the gas is in vertical hydrostatic equilibrium.
\end{enumerate}
These assumptions are likely valid in HOPS-370, as discussed in Section \ref{sec:uncertainties}.
Under these assumptions, we derive a relationship between the ambipolar diffusivity coefficient and observable quantities of
\begin{equation}\label{eq:1}
\eta_\mathrm{AD} \simeq \frac{\delta_r \delta_\phi^2 G^{1/2} C_s^2 R_d^{1/2}(M_\star + M_d)^{1/2} \rho}{B_\phi^2}, 
\end{equation}
where $\delta_r$ and $\delta_\phi$ are scaling factors for the infall and rotational velocities, $G$ is the gravitational constant, $C_s$ is the isothermal sound speed, $R_d$ is the disk radius, $M_\star + M_d$ is the mass of the star+disk system, $\rho$ is the density at the disk-envelope interface and $B_\phi$ is the toroidal (azimuthal) component of the magnetic field strength at the edge of the disk. 
As shown in Appendix \ref{sec:inclinationeffects}, the global magnetic field inclination with respect to the disk rotation axis has little effect on the predicted ambipolar diffusivity coefficient. 
Thus, this prescription should be considered generally valid regardless of the global magnetic field orientation.

To simplify the use of our equation, we select several arbitrary normalization constants to give 

\begin{equation}\label{eq:2}
\begin{split}
\eta_\mathrm{AD} &\simeq 2.5\times10^{17}\,\mathrm{cm^2\,s^{-1}} \left(\delta_r \delta_\phi^2\right) \\
&\times\left(\frac{C_s}{200\,\mathrm{m\,s^{-1}}}\right)^2 \left(\frac{R_d}{100\,\mathrm{au}}\right)^{1/2}\left(\frac{M_\star + M_d}{0.1\,M_\odot}\right)^{1/2} \\
&\times\left(\frac{\rho_d}{1.8\times10^{-15}\,\mathrm{g\,cm^{-3}}}\right) \left(\frac{B_\phi}{20\,\mathrm{mG}}\right)^{-2}.
\end{split}
\end{equation}
In addition, a common normalization used in numerical simulations is to multiply by $4\pi/c^2$, which was used by \citet{Hennebelle2016ApJ...830L...8H} in their derivation to give $\eta_\mathrm{AD}$ in units of seconds.
This normalization produces a relation of
\begin{equation}\label{eq:3}
\begin{split}
\eta_\mathrm{AD} &\simeq 0.0035\,\mathrm{s} \left(\delta_r \delta_\phi^2\right) \\
&\times\left(\frac{C_s}{200\,\mathrm{m\,s^{-1}}}\right)^2 \left(\frac{R_d}{100\,\mathrm{au}}\right)^{1/2}\left(\frac{M_\star + M_d}{0.1\,M_\odot}\right)^{1/2} \\
&\times\left(\frac{\rho_d}{1.8\times10^{-15}\,\mathrm{g\,cm^{-3}}}\right) \left(\frac{B_\phi}{20\,\mathrm{mG}}\right)^{-2},
\end{split}
\end{equation}
which will also be used in later comparisons.

As shown by \citet{Hennebelle2016ApJ...830L...8H}, \citet{Hennebelle2020A&A...635A..67H} and \citet{Commercon2022A&A...658A..52C}, the ratio of the disk radius measured in their numerical simulations ($R_\mathrm{sim}$) to the theoretical disk radius predicted by their ambipolar diffusivity equation ($R_\mathrm{AD}$) was $R_{\mathrm{sim}}/R_\mathrm{AD}\sim1$ (within a factor of $\simeq2-3$) and did not vary considerably over the evolution of the protostellar disks in their simulations. 
Since our main assumptions are essentially the same, this should still hold true even for our new relation. 
This will be explored in more detail in a future paper. 
In the next sections, we describe each of the variables used for our estimate of the ambipolar diffusivity coefficient at the edge of the HOPS-370 protostellar disk for the first time.
This estimation is only possible due to the extensive modeling of HOPS-370 and its surrounding environment from several different observational studies.


\subsection{Previously Estimated Protostar+Disk Properties}\label{sec:diskproperties}
\citet{Tobin2020ApJ...905..162T} derived several important properties of the protostar and disk in HOPS-370. In this section, we describe their extensive molecular line modeling in the context of the relevant values needed for our ambipolar diffusivity coefficient estimation. 

\subsubsection{Protostellar Mass, Disk Mass and Disk Radius}\label{sec:massandradius}
The protostellar masses and disk radii are derived from 12 independent molecular line fits (with a fixed temperature power-law index) using MCMC radiative transfer fitting. 
They found the best fitting protostellar mass to be between ranged between 1.8$\,M_\odot$ and 3.6$\,M_\odot$, with an average protostellar mass of 2.5$\pm$0.2$\,M_\odot$. 
This protostellar mass is the dynamical mass obtained from the Keplerian profile in the line fits.
For the disk radius, the best fits ranged between 70$\,$au and 121$\,$au, with an average radius of 94.4$\pm$12.6$\,$au.
The uncertainties of these average values were determined by using the median-absolute deviation (MAD) of their 12 molecular line fits and scaling them to correspond to one standard deviation of the normal distribution.
We adopt $R_d=94.4\pm12.6\,$au and $M_\star=2.5\pm0.2\,M_\odot$ as the protostellar disk radius and protostellar mass, respectively. 
It is important to note, the $R_d$ used for comparison to $R_\mathrm{AD}$ in the numerical simulations by \citet{Hennebelle2016ApJ...830L...8H} is defined by several conditions using an azimuthally-averaged simulation snapshot: (1) the disk is Keplerian meaning the azimuthal velocity is much greater than the radial velocity, (2) the disk is near hydrostatic equilibrium meaning the azimuthal velocity is much greater than the vertical velocity, (3) the disk is rotationally-supported meaning the rotational energy is larger than the support from thermal pressure by some factor, (4) the disk should be near the equitorial plane, and (5) a density threshold of $n > 10^{9}\,\mathrm{cm^{-3}}$ \citep{Joos2012A&A...543A.128J}.
We have assumed the best-fit gas disk radius is equal to this radius. 
This is further explored in Section \ref{sec:nicilparamcompare}.
The radius of the dust disk also modeled by \citet{Tobin2020ApJ...905..162T}, however the dust is potentially more prone to radial drift and/or optical depth effects \citep[e.g.,][]{Facchini2017A&A...605A..16F}, thus potentially underestimating the actual extent of centrifugal support. 

Additionally, several methods were used by \citet{Tobin2020ApJ...905..162T} to derive the disk mass in HOPS-370. 
First, they used the continuum emission as 1.3$\,$mm, 0.87$\,$mm and 9$\,$mm to derive a value for the disk mass under the assumptions of isothermal and optically thin dust emission.
The disk mass at each wavelength was found to be 0.048$\,M_\odot$ at 0.87$\,$mm, 0.084$\,M_\odot$ at 1.3$\,$mm, and 0.098$\,M_\odot$ at 9$\,$mm. 
They also derive a disk mass from their MCMC radiative transfer fitting of the 0.87$\,$mm dust continuum emission. 
This method resulted in a  disk mass of $0.035^{+0.005}_{-0.003}\,M_\odot$, which is slightly lower than the earlier estimations using the optically thin assumption.
The lower value is likely due to the maximum dust grain size fit of the 0.87$\,$mm emission being 432$\,\mu$m, meaning that the dust in the model will radiate more efficiently than under the assumptions made for the optically thin calculation.
Thus, to be consistent with the dust grain properties later used in our analysis (Section \ref{sec:nicilcompare}), we take the disk mass to be  $M_d=0.035^{+0.005}_{-0.003}\,M_\odot$ for our estimation.
It is important to mention that the uncertainty of the measured disk mass reported by \citet{Tobin2020ApJ...905..162T} are the 1$\sigma$ statistical uncertainties from their MCMC radiative transfer fitting. Thus, these uncertainties likely do not reflect the entire uncertainty of the measured disk mass. 
\citet{Tobin2020ApJ...905..162T} also fit for the disk mass in their 12 molecular line fits. 
However these derived disk masses are highly sensitive to the chosen molecular abundances in the fit, and may not be as reliable. 
This further motivates our choice to use the best-fit disk mass estimated from the dust emission fitting. 

\subsubsection{The Temperature Distribution of the Disk}\label{sec:temperature}

The gas temperature distribution of the HOPS-370 protostellar disk is modeled using a parameterized equation given by
\begin{equation}\label{eq:4}
    T_\mathrm{d}(r) = T_0\left(\frac{r}{1\,\mathrm{au}}\right)^{-q},
\end{equation}
where $T_0$ is the gas temperature at 1$\,\mathrm{au}$ and $q$ is a power-law index, which is fixed to be 0.35 in the 12 molecular line fits by \citet{Tobin2020ApJ...905..162T}. 
The best-fit average value of $T_0$ was found to be $980.0\pm0.6\,\mathrm{K}$, where the errors are also found using the median absolute deviation scaled to one standard deviation of the normal distribution. 
Using the protostellar disk radius of $R_d=94.4\pm12.6\,$au, we find the temperature at the edge of the disk to be $T_d = 199.0\pm9.3\,\mathrm{K}$.\footnote{Uncertainties were propagated using the publicly hosted python package: \textit{asymmetric\_uncertainty} (\citealt{Gobat2022ascl.soft08005G}; \url{https://github.com/cgobat/asymmetric_uncertainty}). This package uses an empirical/analytical function to model the error distributions.}

With this gas temperature, the isothermal sound speed at the disk edge can be estimated by
\begin{equation}\label{eq:5}
    C_s = \left(\frac{k_\mathrm{B} T_d}{\mu_\mathrm{m} m_\mathrm{H}}\right)^{0.5},
\end{equation}
where $k_\mathrm{B}$ is the Boltzmann constant, $\mu_\mathrm{m}=2.37$ is the mean molecular weight for a molecular gas with solar metallicity, and $m_\mathrm{H}$ is the mass of a hydrogen atom. 
The isothermal sound speed is estimated to be $C_s = 833.0\pm19.5\,\mathrm{m\,s^{-1}}$, which is higher than the typically assumed value of $200\,\mathrm{m\,s^{-1}}$ \citep[e.g.,][]{Lee2021ApJ...922...36L,Lee2024ApJ...961L..28L}.

\subsubsection{The Density at the Disk-Envelope Interface}\label{sec:density}

The density at the disk-envelope interface can be estimated via two different approaches. 
The first is by using the best fit values of the disk density profile, while the second is by using the best fit values of the envelope density profile, both modeled by \citet{Tobin2020ApJ...905..162T}.
We initially choose the former approach, since the focus of the study by \citet{Tobin2020ApJ...905..162T} was on the disk, and the observations taken likely resolve out most of the envelope emission.
However, as a comparison, we do explore the latter in Appendix \ref{sec:etaADenvelope}. 
The disk density, which is related to the disk scale height and disk surface density, was modeled using the molecular line emission by \citet{Tobin2020ApJ...905..162T}. 
The disk scale height ($h_d$) is given by 
\begin{equation}\label{eq:6}
    h_\mathrm{d}(r) = \left(\frac{k_\mathrm{B} r^{3} T_d (r)}{G M_\star \mu_\mathrm{m} m_\mathrm{H}}\right)^{0.5},
\end{equation}
where $M_\star$ is the protostellar mass.
The disk surface density ($\Sigma_\mathrm{disk}$) is given by 
\begin{equation}\label{eq:7}
    \Sigma_d(r) = \Sigma_0 \left(\frac{r}{r_c}\right)^{-\gamma} \exp \left[-\left(\frac{r}{r_c}\right)^{(2-\gamma)}\right],
\end{equation}
where $r_c$ is the critical radius of the disk ($r_c = R_d$ was assumed in the molecular line fitting) and $\gamma$ is the surface density power-law index. 
The normalization constant ($\Sigma_0$) is described by
\begin{equation}\label{eq:8}
    \Sigma_0=\frac{(2-\gamma)M_d}{2\pi r_c^2},
\end{equation}
where $M_d$ is the disk mass. 
The radiative transfer modeling of the 12 molecular line fits give an average value of the surface density power-law index to be $\gamma = 0.9\pm0.2$. 
Finally, the disk volume density ($\rho_d$) is expressed as 
\begin{equation}\label{eq:9}
    \rho_d(r)=\frac{\Sigma_d(r)}{\sqrt{2\pi}\ h_d(r)} \exp \left( -\frac{1}{2}\left[\frac{z}{h_d(r)}\right]^2\right),
\end{equation}
where $z$ is the height above the disk midplane and the other parameters are as described before. 
For simplicity, we approximate the density at the midplane ($z=0$), which allows the exponential to go to 1 as the inner terms go to 0.
We are left with a simplified equation of 
\begin{equation}\label{eq:10}
    \rho_d(r)=\frac{\Sigma_d(r)}{\sqrt{2\pi}\ h_d(r)},
\end{equation}
where we can then plug in our known values to calculate the approximate density at the disk edge. 
By plugging in $r = R_d = 94.4\pm12.6\,\mathrm{au}$ and the other parameters previously mentioned, we find a disk scale height of $h_d=16.2\pm3.3\,\mathrm{au}$, a disk surface density of $\Sigma_d=2.2\pm0.9\,\mathrm{g\,cm^{-2}}$, and a disk volume density of $\rho_d=3.7\pm1.7\times10^{-15}\,\mathrm{g\,cm^{-3}}$ at the edge of the disk.

\subsection{Estimating the Magnetic Field Strength at the Edge of the Disk}
\citet{Yen2021ApJ...916...97Y} originally estimated the core-scale plane-of-sky magnetic field strength using 850$\,\mu$m dust polarization legacy observations from the Submillimetre Common-User Bolometer Array (SCUBA) Polarimiter (SCUPOL) on the James Clerk Maxwell Telescope (JCMT). 
A magnetic field strength of $B_\mathrm{pos}=0.54\pm0.25\,\mathrm{mG}$ is derived for HOPS-370 using the Davis-Chandrasekhar-Fermi (DCF) method \citep{Davis1951PhRv...81..890D,Chandrasekhar1953ApJ...118..113C}. 
Updated observations have since been taken using the new SCUBA-2 detector and POL-2 polarimiter (Kao \& Yen et al., in prep.), providing a new and more precise magnetic field strength estimate of $B_\mathrm{pos}=0.50\pm0.13\,\mathrm{mG}$ for HOPS-370. 
In addition to the magnetic field strength, the average core mass and core density were estimated to be $M_c=37.0\pm2.6\,M_\odot$ and $\rho_c=1.9\pm0.2\times10^{-18}\,\mathrm{g\,cm^{-3}}$, respectively, within a core radius of $\sim0.07\,\mathrm{pc}$. 
This is the same radius in which the magnetic field strength was also estimated.
In order to scale this magnetic field strength from the core-scale to the edge of the disk and obtain a value for $B_\phi$, several assumptions need to be made. 

\subsubsection{The Magnetic Field - Density Relation}\label{sec:bnrelation}
The general form of the most commonly cited magnetic field-density ($B$-$n$) relation is written as 
\begin{equation}\label{eq:11}
    B = B_0\left(\frac{n}{n_0}\right)^{\kappa},
\end{equation}
where $B_0$ is the initial magnetic field strength to be scaled, $n$ and $n_0$ are scaled and initial number densities, respectively, and $\kappa$ is the power-law index \citep{Crutcher2010ApJ...725..466C, Crutcher2019FrASS...6...66C,Pattle2023ASPC..534..193P}.
For clouds undergoing spherical collapse with flux-freezing, $\kappa$ is $\sim2/3$ \citep{Mestel1966MNRAS.133..265M}, while collapse models with ambipolar diffusion predict $\kappa$ evolves from 0 at the initial collapse to 0.5 in the later stages \citep{Mouschovias1999ASIC..540..305M}.
Since our primary assumption is that ambipolar diffusion is the main diffusion process, and HOPS-370 is an evolved Class 0 protostar, we take $\kappa=0.5$.  
Thus, the total magnetic field strength can be scaled by 
\begin{equation}\label{eq:12}
    B_\mathrm{tot,d}=B_\mathrm{tot,c}\left(\frac{\rho_\mathrm{d}}{\rho_\mathrm{c}}\right)^{0.5},
\end{equation}
where $\rho_c$ and $\rho_c$ are the volume densities at the core and disk scales, while $B_\mathrm{tot,c}$ and $B_\mathrm{tot,d}$ are the total magnetic field strengths at the core and disk scales, respectively (hereafter, referred to as the C04 method).


\subsubsection{Magnetic Field Strength Scaling, Correction and Estimation}\label{sec:estimatebz}

In order to estimate the magnetic field strength at the edge of the disk ($B_\phi$) as fairly as possible, we first convert the plane-of-sky magnetic field strength ($B_{\mathrm{pos},c}$) to the total magnetic field strength ($B_{\mathrm{tot},c}$) using two different statistical relations for the sake of completeness. 
We first use the relation derived from a sample of observations \citep[][]{Crutcher2004ApJ...600..279C}, given as 
\begin{equation}\label{eq:14}
    B_\mathrm{tot} = \left(\frac{4}{\pi}\right) B_\mathrm{pos},
\end{equation}
which gives a statistical average of the total magnetic field strength. 
Using this relation, we derive a total magnetic field strength of $B_{\mathrm{tot},c}=0.64\pm0.16\,\mathrm{mG}$. 
Additionally, \citet{Liu2021ApJ...919...79L} derive the relation 
\begin{equation}\label{eq:15}
    B_\mathrm{tot} = \sqrt{\frac{3}{2}} B_\mathrm{pos},
\end{equation}
using 3D MHD simulations and radiative transfer calculations to produce synthetic polarization images to find a statistical average of the total magnetic field strength. 
Using this relation, we derive a total magnetic field strength of $B_{\mathrm{tot},c}=0.61\pm0.16\,\mathrm{mG}$. 
These two values are within error, and thus, indistinguishable for our purpose.
We therefore simply adopt the value using the statistical relation from \citet{Crutcher2004ApJ...600..279C} for the remainder of this paper.

We now scale our total core-scale magnetic field strength of $B_{\mathrm{tot},c}=0.64\pm0.16\,\mathrm{mG}$ down to the disk scales using Equation \ref{eq:12}.
We find $B_{\mathrm{tot},d}=28.3\pm9.9\,\mathrm{mG}$ using the C04 method.
Since $B_\phi$ should be the dominant magnetic field component at the edge of the protostellar disk, we assume $B_\phi \sim B_\mathrm{tot,d}$ in our estimations.
This is discussed later in Section \ref{sec:quantityassumptions}.

\begin{deluxetable}{lcr}[t!]
\tabletypesize{\footnotesize}
\tablewidth{\textwidth}
\tablecaption{Overview of parameters used for the estimation of $\eta_\mathrm{AD}$}
\label{tab:parameters}
\tablehead{\colhead{\textbf{Parameter Description}} & \colhead{\textbf{Parameter}} & \colhead{\textbf{Value}}} 
\startdata
\multicolumn{3}{c}{\textbf{Protostar + Disk Properties}} \\
\hline
Protostellar mass & $M_\star$ ($M_\odot$) & $2.5^{+0.2}_{-0.2}$ \\
Disk mass & $M_d$ ($M_\odot$) & $0.035^{+0.005}_{-0.003}$ \\
Disk radius & $R_d$ (au) & $94.4^{+12.6}_{-12.6}$ \\
Critical radius  & $r_c$ (au) & $=R_d$ \\
Temperature at 1 au & $T_\mathrm{0}$ (K) & $980.0^{+0.6}_{-0.6}$ \\
Temperature power-law index & $q$ & $0.35$ \\
Surface density power-law index & $\gamma$ & $0.9^{+0.2}_{-0.2}$ \\
Disk temperature at $R_d$ & $T_d$ (K) & $199.0^{+9.3}_{-9.3}$\\
Disk sound speed at $R_d$ & $C_s$ (m$\,$s$^{-1}$) & $833.0^{+19.5}_{-19.5}$\\
Disk scale height at $R_d$ & $h_d$ (au) & $16.2^{+3.3}_{-3.3}$\\
Disk surface density at $R_d$ & $\Sigma_d$ (g$\,$cm$^{-2}$) & $2.2^{+0.9}_{-0.9}$\\
Disk volume density at $R_d$ & $\rho_d$ ($10^{-15}$g$\,$cm$^{-3}$) & $3.7^{+1.7}_{-1.7}$\\
\hline
\multicolumn{3}{c}{\textbf{Protostellar Core Properties}} \\
\hline
Core mass & $M_c$ ($M_\odot$) & $37.0^{+2.6}_{-2.6}$ \\
Core volume density & $\rho_c$ ($10^{-18}$g$\,$cm$^{-3}$) & $1.9^{+0.2}_{-0.2}$ \\
Plane-of-sky B-field strength & $B_\mathrm{pos,c}$ (mG) & $0.5^{+0.1}_{-0.1}$ \\
\hline
\multicolumn{3}{c}{\textbf{Core to Disk Scale B-Field Properties}} \\
\hline
B-n relation power-law index & $\kappa$ & 0.5 \\
Total Core B-field strength & $B_{\mathrm{tot},c}$ (mG) & $0.6^{+0.2}_{-0.2}$ \\
Total Disk B-field strength & $B_{\mathrm{tot},d}$ (mG) & $28.3^{+9.9}_{-9.8}$ \\
\enddata
\tablerefs{\citet{Tobin2020ApJ...905..162T}, Kao \& Yen et al. (in prep.), this work.}
\end{deluxetable}

\subsection{Scaling Factors for the Infalling and Rotational Velocities}\label{sec:scalingfactors}

Here, we discuss the scaling factors of $\delta_r$ and $\delta_\phi$, which describe the deviations of the infalling and rotational velocities, respectively, from the Keplerian velocity (Equation \ref{eq:A9}). 
As briefly mentioned in Appendix \ref{sec:etaderivation}, recent MHD simulations of protostellar disk formation including ambipolar diffusion find that $u_\phi$ is very close to Keplerian at the disk edge ($\delta_\phi \gtrsim 0.9$), while $u_r$ is significantly less ($\delta_r \lesssim 0.5$) than the Keplerian velocity, possibly by even a factor of a few, less than one order of magnitude \citep{Lee2021A&A...648A.101L}. 
For the deviation of the rotational velocity from Keplerian, we will initially assume $\delta_\phi = 1$ as a conservative estimate, and since the modeling of the HOPS-370 protostellar disk already assumes the rotational velocity structure of the disk is Keplerian.
For the deviation of the infall velocity from Keplerian, it is less straight-forward but we can still make some estimates. 
Recent observations of the young Class I protostar, L1489 IRS, revealed a so-called ``slow'' infall, where the velocity structure of the infalling envelope was modeled to be 2.5 times slower than freefall \citep{Sai2022ApJ...925...12S}. 
If we use the quantities for HOPS-370 and this assumption of $v_\mathrm{inf} = 0.4v_\mathrm{ff}$, we find $\delta_r \sim 0.6$.
For a conservative measure, we initially assume $\delta_r = 0.8$.
Modeling $\delta_r$ in HOPS-370 would provide further constraints on our ambipolar diffusivity coefficient, however, this is currently beyond the scope of this paper and will be left for a future study.
How these two values effect the ambipolar diffusivity coefficient estimation is further explored in Section \ref{sec:nicilparamcompare}.

\section{Results \& Analysis}\label{sec:results}
\subsection{The First Estimation of the Ambipolar Diffusivity Coefficient from Observations}\label{sec:firstestimation}

In the previous sections, we obtained all of the necessary values needed to estimate $\eta_\mathrm{AD}$ for the first time. 
An overview of all the parameters obtained in the previous sections is shown in Table \ref{tab:parameters}.
We make an estimation using the $B_\phi$ derived from the C04 method. 
We plug in the values of
\begin{gather*} 
    \delta_r = 0.8, \\
    \delta_\phi = 1.0, \\
    C_s = 833.0^{+19.5}_{-19.5}\,\mathrm{m\,s^{-1}}, \\
    R_d = 94.4^{+12.6}_{-12.6}\,\mathrm{au}, \\
    M_\star = 2.5^{+0.2}_{-0.2}\,M_\odot, \\
    M_d = 0.035^{+0.005}_{-0.003}\,M_\odot \\
    \rho_d=3.7^{+1.7}_{-1.7}\times10^{-15}\,\mathrm{g\,cm^{-3}} \\
    B_\phi = 28.3^{+9.9}_{-9.8}\,\mathrm{mG}, 
\end{gather*}
into the normalized ambipolar diffusivity coefficient equation (Equations \ref{eq:2} and \ref{eq:3}) to obtain 
\begin{align*}
    \eta_\mathrm{AD} &= 1.7^{+1.5}_{-1.4}\times10^{19}\,\mathrm{cm^{2}\,s^{-1}}\\
    &= 2.4^{+2.1}_{-2.0}\times10^{-1}\,\mathrm{s}.
\end{align*}
As this is the first ever estimation of the ambipolar diffusivity coefficient from observations, there are no other observational values to compare with. 
In the context of comparing to the value of this coefficient using a chemical network, this will be explored in Section \ref{sec:nicilcompare}.

\subsection{The Dimensionless Els\"{a}sser Number for Ambipolar Diffusion}\label{sec:elsassesnumber}

The strength of non-ideal MHD effects are quantified through the dimensionless Els\"{a}sser numbers, which for ambipolar diffusion is given by 
\begin{equation}
    \mathrm{AM} = \frac{v^2_\mathrm{A}}{\eta_\mathrm{AD}\Omega_\mathrm{K}},
\end{equation}
where $v_\mathrm{A}$ is the Alfv\'{e}n speed and $\Omega_\mathrm{K}$ is the Keplerian rotation frequency \citep[e.g.,][]{Wurster2021MNRAS.501.5873W,Cui2021MNRAS.507.1106C}.
The Alfv\'{e}n speed is defined as 
\begin{equation}
    v_\mathrm{A} = \sqrt{\frac{B^2}{4\pi\rho}},
\end{equation}
which describes the speed of an MHD wave permeating through a dense medium.
Likewise, the Keplerian rotation frequency is defined as 
\begin{equation}
    \Omega_\mathrm{K} = \sqrt{\frac{GM_\star}{r^3}}.
\end{equation}
Typically, $\mathrm{AM} \gg 1$ represents strong coupling between the magnetic field and the neutral gas, while $\mathrm{AM} \lesssim 1$ indicates strong magnetic diffusion \citep[e.g.,][]{Wurster2021MNRAS.501.5873W,Commercon2022A&A...658A..52C}.
We estimate the dimensionless Els\"{a}sser number for ambipolar diffusion to be $\mathrm{AM = 1.7\pm1.0}$.
This shows we are likely in the regime of stronger magnetic diffusion and indicates the importance of ambipolar diffusion in the evolution of the HOPS-370 protostellar disk. 

\subsection{Comparing with the Non-Ideal MHD Coefficient and Ionisation Library (NICIL)}\label{sec:nicilcompare}

The Non-Ideal MHD Coefficient and Ionisation Library (NICIL)\footnote{\url{https://bitbucket.org/jameswurster/nicil/src/master/}} is a code to calculate the diffusion coefficients for ambipolar diffusion ($\eta_\mathrm{AD}$), Ohmic dissipation ($\eta_\mathrm{OD}$) and Hall effect ($\eta_\mathrm{HE}$) for MHD simulations using a chemical network \citep{NICIL2016PASA...33...41W,Wurster2021MNRAS.501.5873W}.
We aim to investigate whether our ambipolar diffusivity coefficient is consistent with one calculated by NICIL.
NICIL allows for estimating these coefficients for different input parameters, such as density, temperature and magnetic field strength.
Additionally, parameters for the dust grain size distribution and cosmic-ray ionization rate can be modified.
First, we describe the initial parameters used for several NICIL runs (Section \ref{sec:nicilsetup}).
We attempt to emulate the conditions at the edge of the disk as closely as possible by using the derived disk parameters and several different assumptions for the magnetic field strength and the cosmic-ray ionization rate (Section \ref{sec:nicilinitialcompare}). 
We then explore several of the assumptions made during our estimation of the ambipolar diffusivity coefficient to see how they affect the value and its consistency with NICIL (Section \ref{sec:nicilparamcompare}).
There are two files that we modify for these different runs in NICIL: \texttt{nicil.F90} and \texttt{nicil{\_}ex{\_}eta.F90}.
We assume a barotropic equation of state for all runs, since this is the same assumption used in the numerical simulations by \citet{Hennebelle2016ApJ...830L...8H}.

\begin{figure*}[th!]
  \centering
  \includegraphics[width=\textwidth]{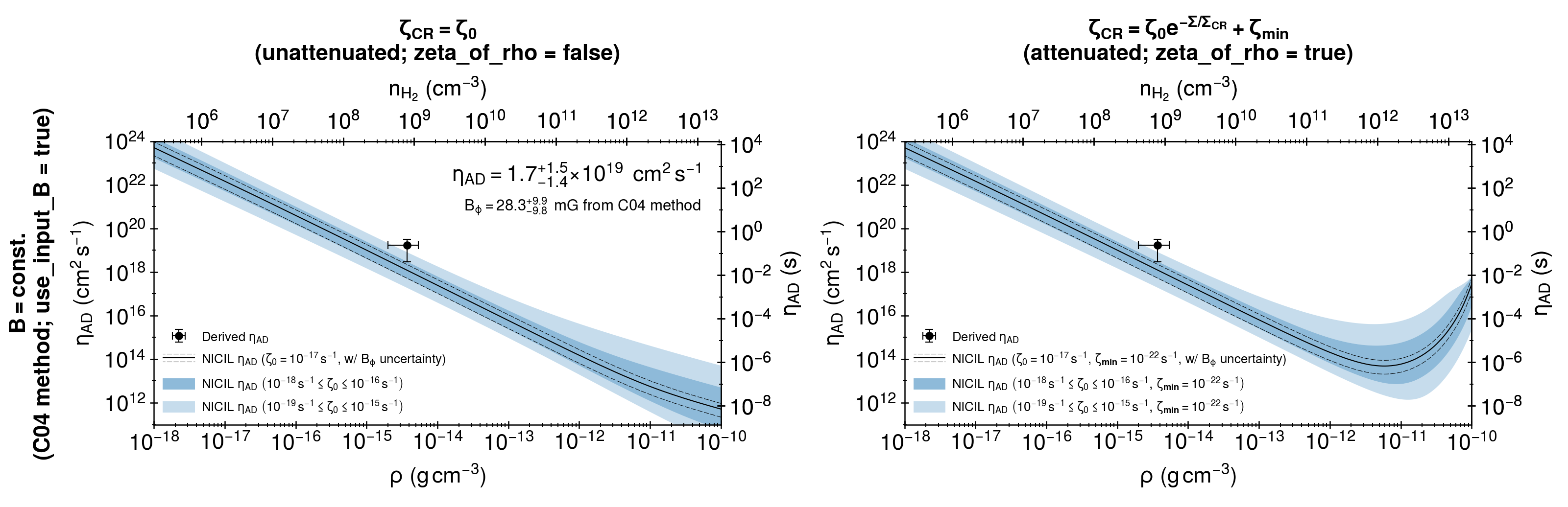}
  \caption{Comparison between our derived value of $\eta_\mathrm{AD}$ and the NICIL calculated values of $\eta_\mathrm{AD}$ assuming a barotropic equation of state and a constant magnetic field strength (C04 method).
  Our derived $\eta_\mathrm{AD}$ is marked by the black circle and also printed in the top right of each plot, along with the values used for the estimation just below. 
  \textbf{(Left Column)} Uses a constant (unattenuated) cosmic-ray ionization rate for the NICIL calculation. 
  \textbf{(Right Column)} Uses a varied (attenuated) cosmic-ray ionization rate for the NICIL calculation. 
  The dashed black lines indicate the NICIL calculated $\eta_\mathrm{AD}$ based on the magnetic field strength uncertainties.
  The shaded blue areas represent $\eta_\mathrm{AD}$ calculated by NICIL for different ranges of $\zeta_\mathrm{0}$ between $10^{-16} - 10^{-18}\,\mathrm{s^{-1}}$ (darker shade) and $10^{-15} - 10^{-19}\,\mathrm{s^{-1}}$ (lighter shade).
  The mass volume density and H$_2$ number density are related by $\rho = m_\mathrm{H} \mu_\mathrm{H_2} n_\mathrm{H_2}$, where $m_\mathrm{H}$ is the mass of the hydrogen atom and $\mu_{H_2}$ is the mean molecular weight per molecular hydrogen ($\mu_\mathrm{H_2}=2.8$). 
  }
  \label{fig:nicilinitialcomparison}
\end{figure*}

\subsubsection{Parameter Setup}\label{sec:nicilsetup}

We first describe several modifications made to the \texttt{nicil{\_}ex{\_}eta.F90} test script.
This file contains the input parameters for the temperature, density and magnetic field strength.
We compute the barotropic equation of state over the default temperature range of $10\,$K to $2\times10^{5}\,$K and density range of $10^{-22}\,\mathrm{g\,cm^{-3}}$ to $10^{0.5}\,\mathrm{g\,cm^{-3}}$.
For the magnetic field, we employ a constant (\texttt{use{\_}input{\_}B = .true.}) magnetic field strength using a value of 28.3$\,$mG (C04 method), which we estimated at the edge of the disk.
We also run using the upper and lower errors on the magnetic field as the constant values to estimate an approximate error range on the NICIL ambipolar diffusivity coefficient. 
Additionally, NICIL has the option to vary the magnetic field using the function $B = 1.34\times10^{-7}\sqrt{n_n}\,\mathrm{G}$ (\texttt{use{\_}input{\_}B = .false.}).
However, this magnetic field strength comes from different underlying assumptions than what we use and severely underestimates the magnetic field strengths compared to what we find, so we do not compare with this case.

The dust grain and cosmic-ray ionization properties are then adjusted in the \texttt{nicil.F90} main script. 
We use the default gas-to-dust ratio of 100 and set the number of grain size bins to 32.
\citet{Tobin2020ApJ...905..162T} derive power-law slope of the grain distribution to be $p=-2.63$ and the maximum grain size to be $a_\mathrm{max} = 432\,\mathrm{\mu m}$, while assuming the same minimum dust grain size of $a_\mathrm{min} = 0.005\,\mathrm{\mu m}$ used in their fitting.
Thus, we set these parameters in NICIL accordingly.
The cosmic-ray ionization rate in HOPS-370 is unknown, however the typical ISM value is usually quoted to be $\zeta_\mathrm{CR} = 10^{-17}\,\mathrm{s^{-1}}$ \citep[e.g.,][]{Caselli1998ApJ...499..234C,McElroy2013A&A...550A..36M}. 
We initially set a constant cosmic-ray ionization rate (\texttt{zeta{\_}of{\_}rho = .false.}) of $\zeta_\mathrm{CR} = \zeta_\mathrm{0} = 10^{-17}\,\mathrm{s^{-1}}$ in the script.
However, we also vary the cosmic-ray ionization rates between $10^{-19}\,\mathrm{s^{-1}} < \zeta_\mathrm{0} < 10^{-15}\,\mathrm{s^{-1}}$ as another approximate ``error'' range.
This should be a typical range in dense molecular clouds inferred from chemical analyses \citep[][]{Caselli1998ApJ...499..234C}. 
In addition, we also run using a varied cosmic-ray ionization rate (\texttt{zeta{\_}of{\_}rho = .true.}), which mimics attenuated cosmic-rays via the relation of $\zeta_\mathrm{CR}=\zeta_\mathrm{0} e^{-\Sigma / \Sigma_\mathrm{CR}} + \zeta_\mathrm{min}$.
In this case, we set $\zeta_\mathrm{0} = 10^{-17}\,\mathrm{s^{-1}}$  and $\zeta_\mathrm{min} = 10^{-22}\,\mathrm{s^{-1}}$ (default value) in the script. 
The the gas surface (column) density ($\Sigma$) is directly calculated from several parameters when running the code. 
The cosmic-ray attenuation depth ($\Sigma_\mathrm{CR}$) is a constant and kept at the default value of 96$\,\mathrm{g\,cm^{-2}}$.
We also vary the cosmic-ray ionization rates between $10^{-19}\,\mathrm{s^{-1}} < \zeta_\mathrm{0} < 10^{-15}\,\mathrm{s^{-1}}$ in this case as well.
We set the mean molecular weight to be $\mu=2.37$ for consistency.
All other parameters in both scripts are kept as the default values.

\subsubsection{Initial Comparison}\label{sec:nicilinitialcompare}

We run NICIL using the aforementioned parameters and compare to derived values of $\eta_\mathrm{AD}$ in Figure \ref{fig:nicilinitialcomparison}. 
The columns correspond to the two different cosmic-ray ionization rate assumptions in NICIL.
The derived value of $\eta_\mathrm{AD}$ is shown in the top right corner of the left panel plot along with the magnetic field strength from the C04 method. 
We describe each case in more detail below. 
Our $\eta_\mathrm{AD}$ result surprisingly consistent to the $\eta_\mathrm{AD}$ calculated by NICIL. 

\textbf{Constant $\mathbf{B}$ \&  $\mathbf{\zeta}_\mathrm{\mathbf{CR}}$:}
The left panel of Figure \ref{fig:nicilinitialcomparison} use a constant magnetic field strength (\texttt{use{\_}input{\_}B = .true.}) of $B=28.3^{+9.9}_{-9.8}\,\mathrm{mG}$ from the C04 method (left panel). 
In addition, we use a constant (unattenuated) cosmic-ray ionization rate (\texttt{zeta{\_}of{\_}rho = .false.}).
As mentioned in the previous section, we run for the derived magnetic field strength of $28.3\,\mathrm{mG}$, and then perform subsequent runs using the upper/lower errors on the magnetic field strength (dashed black lines).
We assumed the cosmic-ray ionization rate to be $10^{-17}\,\mathrm{s^{-1}}$ for the previous three calculations, but also varied it between $10^{-18}\,\mathrm{s^{-1}} \leq \zeta_0 \leq 10^{-16}\,\mathrm{s^{-1}}$ (darker-blue shaded area) and $10^{-19}\,\mathrm{s^{-1}} \leq \zeta_0 \leq 10^{-15}\,\mathrm{s^{-1}}$ (lighter-blue shaded area) assuming the magnetic field strength of $28.3\,\mathrm{mG}$.
The derived $\eta_\mathrm{AD}$ using the C04 magnetic field strength is surprisingly consistent with the results from NICIL. 
If the infall velocity is much smaller than Keplerian rotation, then both values would become more consistent (Section \ref{sec:nicilparamcompare}).
The cosmic-ray ionization rates have more of an affect on the predicted ambipolar diffusivity coefficient from NICIL compared to the error on our magnetic field strength.
Higher cosmic-ray ionization rates correspond to smaller $\eta_\mathrm{AD}$ values and vice-versa. 
The choice of the chemical network would also impact the estimation by NICIL, and thus our overall comparison. This is discussed more in Section \ref{sec:nicilassumptions}.

\begin{figure*}[ht!]
  \centering
  \includegraphics[width=\textwidth]{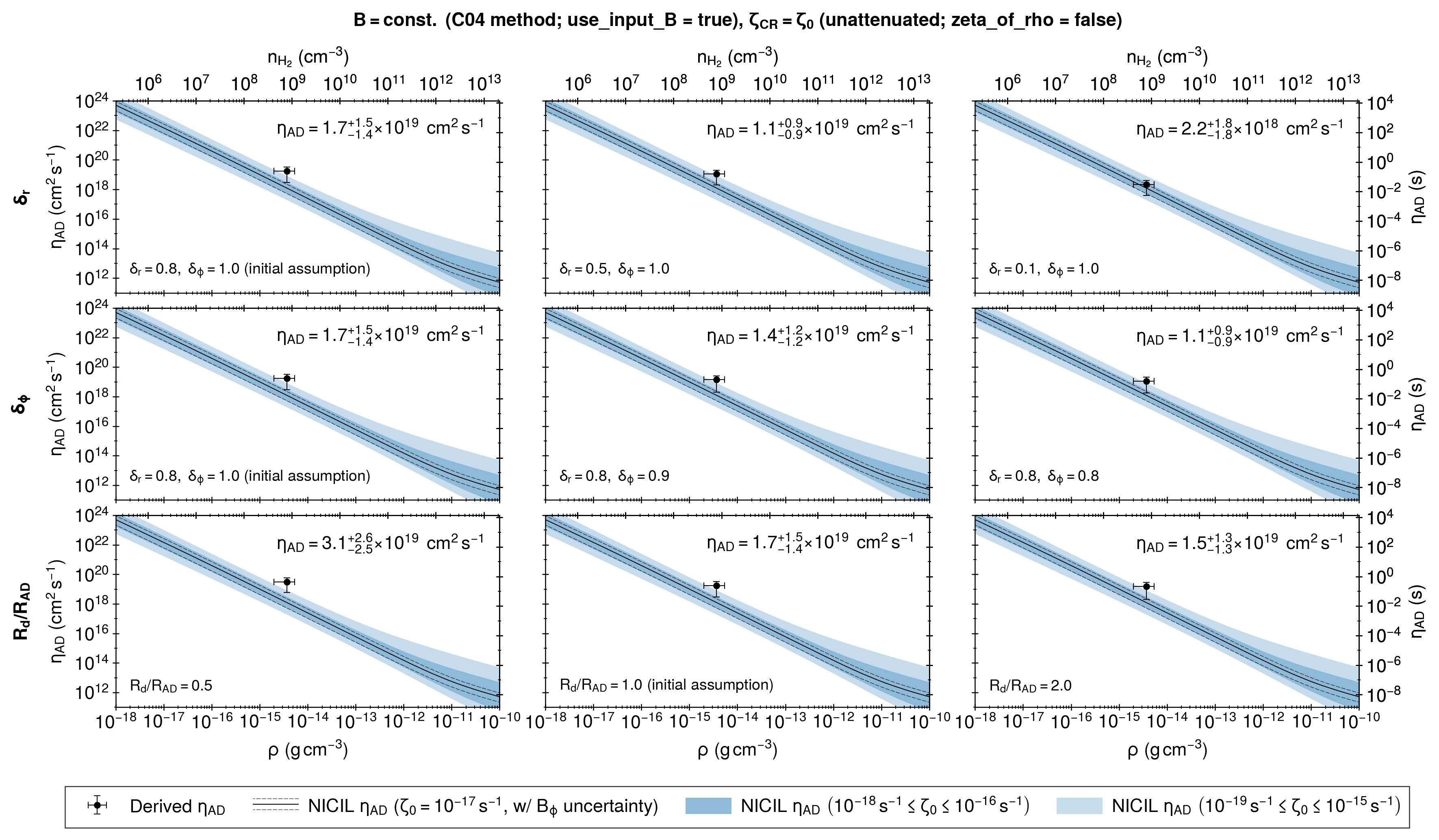}
  \caption{Comparison between the effect of different values of $\delta_r$ (top row), $\delta_\phi$ (middle row) and $R_d / R_\mathrm{AD}$ (bottom row) on our derived $\eta_\mathrm{AD}$ and the NICIL calculated values of $\eta_\mathrm{AD}$ assuming a barotropic equation of state, constant magnetic field strength from the C04 method and constant (unattenuated) cosmic-ray ionization rate. As in Figure 1, our derived $\eta_\mathrm{AD}$ is marked by the black circle and also printed in the top right of each plot, while the black lines and shaded blue areas also have the same meanings as in Figure \ref{fig:nicilinitialcomparison}. 
  }
  \label{fig:fig:nicilC04varied}
\end{figure*}

\textbf{Constant $\mathbf{B}$ \&  Varied $\mathbf{\zeta}_\mathrm{\mathbf{CR}}$:}
The right panel of Figure \ref{fig:nicilinitialcomparison} uses the same magnetic field parameters as the left, however, the cosmic-ray ionization rate is attenuated (\texttt{zeta{\_}of{\_}rho = .true.}). 
We do this in order to understand what effect this has on our derived ambipolar diffusivity coefficient.
As we can see, the attenuated cosmic-rays only affect the very high densities $\gtrsim 10^{-12}\,\mathrm{g\,cm^{-3}}$, where the ambipolar diffusivity coefficient begins to increase. 
It does not affect the density regime in which our ambipolar diffusivity coefficient is estimated. 
Thus, our derived $\eta_\mathrm{AD}$ is still consistent with these results from NICIL. 

\subsubsection{Varied Parameters}\label{sec:nicilparamcompare}

There are several parameter assumptions made in our estimation using the ambipolar diffusivity coefficient equation that could vary the resulting $\eta_\mathrm{AD}$.
First, our initial calculation assumes $\delta_r=0.8$ and $\delta_\phi=1.0$. 
Additionally, the results from \citet{Hennebelle2016ApJ...830L...8H} and \citet{Hennebelle2020A&A...635A..67H} indicate that the ratio of the disk radius from their simulations to their predicted disk radius from their analytical equation, $R_d / R_\mathrm{AD}$, could vary between 0.5 to 2, particularly for lower mass cores.
Therefore, we vary at each parameter individually, while the others are kept at their initially assumed values, to see the magnitude in difference for each. 
We compare to our NICIL run using constant $B$ from the C04 method and constant $\zeta_\mathrm{CR}$ (Figure \ref{fig:fig:nicilC04varied}, left panel).

For $\delta_r$, we compare values of 0.8, 0.5 and 0.1 (Figure \ref{fig:fig:nicilC04varied}, top row).
As mentioned previously, $\delta_r$ could be a factor of a few lower than the Keplerian velocity in simulations, and was found to be $\sim0.6$ in previous observations of the Class I protostar, L1489 IRS.
As this value could vary quite considerably depending on the environment, it will have more of an impact on our estimated ambipolar diffusivity coefficient than $\delta_\phi$, even though $\eta_\mathrm{AD}$ scales as $\sim \delta_\phi^2$.
In the case where $\delta_r = 0.1$, it is much more consistent with NICIL as the estimated ambipolar diffusivity coefficient is about an order of magnitude lower, potentially hinting at the possibility of slow infall in HOPS-370.

For $\delta_\phi$, we compare values of 1.0, 0.9 and 0.8 (Figure \ref{fig:fig:nicilC04varied}, middle row).
Since $\delta_\phi$ should be $\gtrsim 0.9$, it will not have too much affect on our derived ambipolar diffusivity coefficient, which are all within error in these cases. 
We only demonstrate $\delta_\phi=0.8$ as a more extreme scenario, but still the effect is less than $\delta_r$ due to these constraints. 
It would still be interesting to try to estimate if there is any deviation from Keplerian rotation in the rotational velocity structure at the edge of the disk, as it still lowers the value, if only even a little. 

For $R_d / R_\mathrm{AD}$, we compare values of 0.5, 1.0 and 2.0 (Figure \ref{fig:fig:nicilC04varied}, bottom row).
As previously mentioned, the ratio of the actual disk radius to the predicted disk radius due to ambipolar diffusion from Equation \ref{eq:1} could vary between 0.5 to 2, particularly for lower mass cores \citep{Hennebelle2016ApJ...830L...8H, Hennebelle2020A&A...635A..67H}.
Although HOPS-370 is considered to be a more intermediate mass Class 0/I, we would still like to investigate how the value varies between these two extreme cases.
The $R_d/R_\mathrm{AD}$ factor also doesn't really affect the calculation of $\eta_\mathrm{AD}$ too much, which is similar to $\delta_\phi$.
$R_d/R_\mathrm{AD}$ is slightly more consistent with NICIL when $R_d/R_\mathrm{AD} > 1$, while the right panel of Figure 2 in \citet{Hennebelle2016ApJ...830L...8H} shows consistently lower $R_d/R_\mathrm{AD} < 1$ for protostars $M_\star+M_d<5\,M_\odot$.
This should be investigated in numerical simulations for a mass range around the HOPS-370 protostar+disk mass as the spread can become quite noticeable when zooming in on very low-mass simulations ($M_\star+M_d<0.5\,M_\odot$) in the left panel of Figure 2 in \citet{Hennebelle2016ApJ...830L...8H}.

\section{Discussion} \label{sec:discussion}

\begin{figure*}[ht!]
  \centering
  \includegraphics[width=\textwidth]{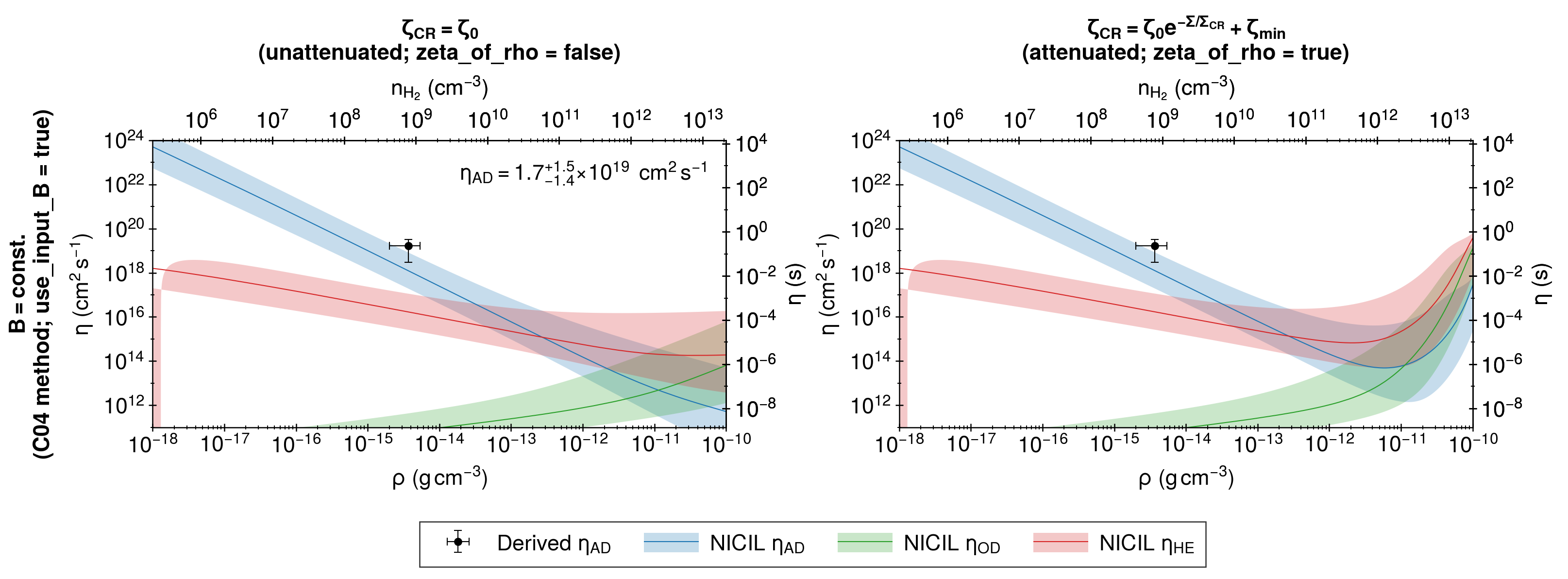}
  \caption{Relative comparison of the diffusivity coefficients for ambipolar diffusion ($\eta_\mathrm{AD}$), Ohmic dissipation ($\eta_\mathrm{OD}$) and the Hall effect ($\eta_\mathrm{HE}$) calculated by NICIL assuming a barotropic equation of state, constant magnetic field strength and cosmic-ray ionization rate.
  The symbols and labels have the same meaning as in Figure \ref{fig:nicilinitialcomparison}, except we only show the lighter shaded blue area for $\zeta_\mathrm{0}$ between $10^{-15} - 10^{-19}\,\mathrm{s^{-1}}$ for each coefficient.
  Our derived value shows that HOPS-370 lies in an ambipolar diffusion dominated region. 
  }
  \label{fig:niciletacomparison}
\end{figure*}

\subsection{Validity of Assumptions}\label{sec:uncertainties}

Several assumptions are made in the derivation, estimation and comparison to theoretical values of the ambipolar diffusivity coefficient. 
In this section, we explore these assumptions in detail and discuss how they could effect our results.

\subsubsection{Derivation of the Ambipolar Diffusivity Coefficient Equation}\label{sec:derivationassumptions}

In Section \ref{sec:methods}, we have listed several assumptions in the derivation of Equation \ref{eq:1}. 
The first is that the main diffusion process is ambipolar diffusion.
There are many factors shown to alleviate the effects of magnetic braking to form large, protostellar disks in MHD simulations. 
These mainly include non-ideal MHD \citep[e.g.,][]{Li2011ApJ...738..180L,Dapp2012A&A...541A..35D,Tsukamoto2015MNRAS.452..278T,Wurster2016MNRAS.457.1037W,Tsukamoto2017PASJ...69...95T,Wurster2019MNRAS.489.1719W,Zhao2020MNRAS.492.3375Z,Wurster2021MNRAS.507.2354W}, misalignment between the magnetic field and rotation axis \citep[e.g.,][]{Hennebelle2009A&A...506L..29H,Li2013ApJ...774...82L,Tsukamoto2018ApJ...868...22T,Hirano2020ApJ...898..118H} and turbulence \citep[e.g.,][]{Seifried2013MNRAS.432.3320S,Li2014ApJ...793..130L,Seifried2015MNRAS.446.2776S}. 
From an observational standpoint, there are several key results to consider.
Magnetic field orientations in low-mass protostars indicate that the field orientation is preferentially randomly aligned with the rotation axis \citep[e.g.,][]{Hull2013ApJ...768..159H,Yen2021ApJ...907...33Y}
However, recent results show no apparent correlation between the misalignment angle of the magnetic field and apparent disk size measured from the dust continuum \citep{Yen2021ApJ...916...97Y}. 
\citet{Yen2021ApJ...916...97Y} also conclude that the turbulence measured from the non-thermal linewidth at core-scale does not correlate with the apparent disk size either. 
Observations of angular momentum profiles in protostellar envelopes do imply that there is some turbulence present \citep{Pineda2019ApJ...882..103P,Gaudel2020A&A...637A..92G,Sai2023ApJ...944..222S}, although the level of turbulence is not directly quantified.
Thus, non-ideal MHD likely plays an important role in protostellar disk formation.

As far as which non-ideal MHD effect (ambipolar diffusion, Ohmic dissipation or the Hall effect) is most important overall, simulations show ambipolar diffusion is an efficient process in parts of the disk and envelope that can regulate the properties of the disk \citep[e.g.,][]{Tsukamoto2023ASPC..534..317T}.
Ohmic dissipation is only efficient at high densities and likely does not play much of a role in the envelope itself \citep[e.g.,][]{Marchand2016A&A...592A..18M,Wurster2018MNRAS.475.1859W,Wurster2021MNRAS.501.5873W}.
The Hall effect has been shown to effectively disappear shortly after the formation of the protostellar disk \citep{Zhao2020SSRv..216...43Z,Lee2021ApJ...922...36L}.
Therefore, ambipolar diffusion may be the most important non-ideal MHD effect, especially when the protostar + disk system becomes more evolved.
This is where comparing directly with non-ideal MHD simulations would help us to understand the non-ideal MHD effects more deeply. 

Since NICIL does also calculate the non-ideal MHD coefficients for Ohmic dissipation ($\eta_\mathrm{OD}$) and the Hall effect ($\eta_\mathrm{HE}$), it is interesting to compare them to the ambipolar diffusivity coefficient. 
We show a comparison between each of the diffusivity coefficients from our NICIL runs in Figure \ref{fig:niciletacomparison}. 
We see that our derived $\eta_\mathrm{AD}$ is clearly in an ambipolar diffusion dominated density regime, under our assumptions made for our NICIL runs. 
This is in favor of the first assumption stated to derive Equation \ref{eq:1}, that ambipolar diffusion is the main diffusion process. 
The coefficient for Ohmic dissipation starts to become prominent towards the highest density regimes, which is consistent with previous findings \citep[e.g.,][]{Marchand2016A&A...592A..18M,Wurster2018MNRAS.475.1859W,Wurster2021MNRAS.501.5873W}.
The Hall effect seems to be dominant at intermediate density regimes, though it still shows some contribution in the density regime where our ambipolar diffusion value is calculated. 
We also note that in our NICIL run in the bottom left panel, the Hall coefficient becomes negative at very low densities approaching $10^{-18}\,\mathrm{g\,cm^{-3}}$ in the case of a low cosmic-ray ionization rate of $\zeta_0 = 10^{-19}\,\mathrm{s^{-1}}$.

The cosmic-ray ionization rate will impact the efficiency of non-ideal MHD effect \citep[e.g.,][]{Wurster2018MNRAS.476.2063W,Kuffmeier2020A&A...639A..86K}, and thus should be studied in the environment of HOPS-370 to further constrain the comparisons with NICIL.
The cosmic-ray ionization rate in the inner envelope of Class 0 protostar, B335, was previously found to be enhanced ($\zeta_\mathrm{CR} \sim 10^{-14}\,\mathrm{s^{-1}}$), which could explain the extremely small ($<10\,\mathrm{au}$) inferred protostellar disk \citep{Cabedo2023A&A...669A..90C}. 
Additionally, a new large scale study probing the NGC 1333 region of the Perseus also finds an enhanced cosmic-ray ionization rate ($\zeta_\mathrm{CR} \gtrsim 10^{-16.5}\,\mathrm{s^{-1}}$) across the molecular cloud, which is consistent with the small ($<50\,\mathrm{au}$) disks in that region \citep{Pineda2024arXiv240216202P}.

Future studies also probing the ion-neutral drift in HOPS-370 could help to further understand the role of ambipolar diffusion in this source.
\citet{Yen2018A&A...615A..58Y} tried to constrain the ion-neutral velocity drift in the young Class 0 protostar, B335, however, only an upper limit was obtained. 
This could be due to B335 being too young, as some simulations have shown this velocity drift could be more observable in more evolved Class 0/I protostars \citep[e.g.,][]{Tsukamoto2020ApJ...896..158T}.
Since HOPS-370 is more evolved, it could be an ideal target for this kind of study in the future. 

Next, the angular momentum is counteracted by magnetic braking resulting in the advection and braking timescales to be of the same order. 
The equations for the advection and braking timescales are given by Equations \ref{eq:A1} and Equations \ref{eq:A2}, respectively.
We make an estimation of the advection timescale under the same assumption used for our ambipolar diffusivity estimate, where $u_r = 0.8 v_\mathrm{kep}$, giving us a value of $\tau_\mathrm{adv}\sim3.6\times10^9\,\mathrm{s}$, which is also a lower limit. 
For the braking timescale, since we do not directly know the poloidal component ($B_z$) of the magnetic field strength, we make an approximation of $B_z B_\phi \approx B_\mathrm{tot}^2$ and use our value derived from the C04 method.
This results in a lower limit of the braking timescale of $\tau_\mathrm{br}\sim5.4\times10^8\,\mathrm{s}$.
These values are within one order of magnitude difference, and show that this assumption can hold in HOPS-370.
We note that we use this exact assumption to estimate $B_z$ in Section \ref{sec:quantityassumptions}, therefore, using that value here presents a circular argument which is why we simply estimate the lower limits for $\tau_\mathrm{adv}$ and $\tau_\mathrm{br}$.
Further modeling of the infall velocity structure and magnetic field components ($B_r$, $B_z$, $B_\phi$) would be necessary to confirm.

Then, the toroidal field generated by differential rotation is offset by the ambipolar diffusion in the vertical direction resulting in the Faraday induction and vertical diffusion timescales to be of the same order. 
The equations for the Faraday induction and vertical diffusion timescales are given by Equations \ref{eq:A3} and Equations \ref{eq:A4}, respectively.
We can obtain a lower limit approximation of the Faraday induction timescales by assuming $B_z \sim B_\phi \sim B_\mathrm{tot}$. 
This gives a value of $\tau_\mathrm{far}\sim4.9\times10^8\,\mathrm{s}$.
The vertical ambipolar diffusion timescales would need to use our derived value, thus we check for self-consistency.
We find $\tau_\mathrm{diff}\sim3.4\times10^9\,\mathrm{s}$.
Since varying some of the parameters in Section \ref{sec:nicilparamcompare} lower the value of $\eta_\mathrm{AD}$, $\tau_\mathrm{diff}$ could also be considered as a lower limit.
Both values are within one order of magnitude difference, showing that this assumption can hold in HOPS-370. 
Again, we do not use the $B_z$ in Section \ref{sec:quantityassumptions} to avoid any circular arguments. 

Additionally, the infalling and rotational velocities of the gas near the disk edge both scale with the Keplerian velocity.
The disk radius derived for HOPS-370 from the radiative transfer modeling indicates that the rotational velocity ($u_\phi$) should be Keplerian in nature. 
As for the infalling velocity, further modeling needs to be done to see how much $u_r$ deviates from Keplerian at the disk edge. 
For now, the assumption of the rotational velocity holds, while the assumption for the infalling velocity should to be further modeled.

Lastly, the gas near the disk edge has Keplerian velocity and is in vertical hydrostatic equilibrium.
Again, the gas disk in HOPS-370 is clearly resolved by the observations by \citet{Tobin2020ApJ...905..162T} and the best-fit disk parameters were found by fitting with radiative transfer models assuming Keplerian rotation and hydrostatic equilibrium. 
Many previous studies have also clearly resolved Keplerian rotating disks in young Class 0 and Class I protostars \citep[e.g.,][]{Tobin2012Natur.492...83T,Murillo2013A&A...560A.103M,Yen2014ApJ...793....1Y,Yen2017ApJ...834..178Y,Ohashi2023ApJ...951....8O}. 
Even if the assumptions in the fitting are wrong, they are the same assumptions we use and we are still using a ``best-fit'' value, which indicates this model does provide a good fit to the data.
Therefore, the values properties derived for the HOPS-370 protostellar disk clearly should satisfy both assumptions.

\subsubsection{Quantities and Relations used for the Ambipolar Diffusivity Coefficient Estimation}\label{sec:quantityassumptions}

Arguably the most important assumption in our estimation of $\eta_\mathrm{AD}$ is how the envelope scales from core-scale down to the edge of the protostellar disk. 
As previously stated, early theoretical works predict $\kappa$ in Equation \ref{eq:4} to be $\sim2/3$ for clouds undergoing spherical collapse with flux-freezing \citep{Mestel1966MNRAS.133..265M}, while $\kappa\sim0.5$ for a collapsing cloud with ambipolar diffusion \citep{Mouschovias1999ASIC..540..305M}.
This was the basis of our initial assumptions, however, it is not so straight forward. 
The recent review by \citet{Pattle2023ASPC..534..193P} shows $\kappa$ derived from observations of molecular clouds can vary quite a bit, possibly due to different environmental factors. 
These observations probe the large scale molecular clouds, filaments and cores whose magnetic field imprint could be inherently different than the magnetic fields near a protostellar disk. 
Additionally, a magnetic field - density relation recently derived by \citet{Lee2024ApJ...961L..28L} for inside a collapsing, protostellar envelope is explored in Appendix \ref{sec:etaADL24}. 
The magnetic field strength derived from this relation is compatible with our estimates, however, needs to be further investigated due to discrepancies in the model presumptions.

Observationally, \citet{Yen2023ApJ...942...32Y} recently derived a magnetic field - density relation from the core to inner envelope scale in the young, Class 0 protostar HH 211. 
They find $\kappa\sim0.36$, which fits into the assumption that ambipolar diffusion is playing a role to partially decoupled the magnetic field from the neutral matter.
Their inner envelope magnetic field strength was derived using a force-balance equation \citep{Koch2012ApJ...747...79K}, rather than the DCF method. 
The core-scale magnetic field strength estimated by Kao \& Yen et al. (in prep.) was derived using the DCF method, which has several uncertainties associated with it due to the assumptions of equipartition, isotropic turbulence, projected polarization angle on the plane-of-sky, and more \citep[e.g.,][]{Liu2021ApJ...919...79L, Liu2022ApJ...925...30L,Chen2022MNRAS.514.1575C,Liu2022FrASS...9.3556L,Myers2023arXiv231209330M}.
These uncertainties may cause the DCF estimate to overestimate the magnetic field strength, which would impact our ambipolar diffusivity coefficient estimation. 
In the best case scenario, future observations to derive the inner envelope strength near the disk in HOPS-370 could alleviate the need to even use a magnetic field - density relation. 
Otherwise, if the magnetic field strength cannot be derived close enough to the disk edge, it can still be estimated in the envelope to derive a magnetic field - density relation, where the magnetic field strength could further be scaled down to the edge of the disk.

We have also assumed that $B_\mathrm{tot} \approx B_\phi$, and that $B_\phi$ is the dominant component of the magnetic field at the edge of the protostellar disk.
Our value for $B_\mathrm{tot}$ at the core-scale is a statistical average based on a large sample of observations, which may or may not necessarily be applied to only a single source. 
This is, however, the only current way we obtain a total magnetic field strength from the plane-of-sky magnetic field component and should be investigated further. 
To see whether $B_\phi$ is really dominant in our case, we estimate $B_z = 4.2\pm2.7\,\mathrm{mG}$ using Equation \ref{eq:A5}. 
This shows that $B_\phi$ is the dominant component in our case, and thus is a reasonable assumption in our ambipolar diffusivity coefficient estimation.

\subsubsection{Comparison with NICIL and Input Values}\label{sec:nicilassumptions}

While the cosmic-ray ionization rate and dust grain properties needed for NICIL are not inherently part of our derived ambipolar diffusion equation, they still play a role in the efficiency of non-ideal MHD diffusivities \citep[e.g.,][]{Zhao2016MNRAS.460.2050Z,Dzyurkevic2017A&A...603A.105D,Wurster2018MNRAS.476.2063W,Zhao2018MNRAS.478.2723Z,Kuffmeier2020A&A...639A..86K,Guillet2020A&A...643A..17G,Zhao2021MNRAS.505.5142Z,Kobayashi2023MNRAS.521.2661K}.
Several studies have shown that disk formation is suppressed in the presence cosmic-ray ionization rates higher than the canonical value of $10^{-17}\,\mathrm{s^{-1}}$ in dense cores \citep[e.g.,][]{Zhao2016MNRAS.460.2050Z,Wurster2018MNRAS.476.2063W,Kuffmeier2020A&A...639A..86K}.
Large numbers of small dust grains can also influence the ionization degree, and thus the non-ideal MHD diffusivities \citep[e.g.,][]{Zhao2016MNRAS.460.2050Z,Dzyurkevic2017A&A...603A.105D,Zhao2018MNRAS.478.2723Z,Koga2019MNRAS.484.2119K,Marchand2020ApJ...900..180M}.
\citet{Tobin2020ApJ...905..162T} do constrain the maximum grain size, while the minimum grain size is set as a fixed parameter in their model fitting. 
We did explore how much the minimum grain size affects the calculated $\eta_\mathrm{AD}$ from NICIL by re-running our constant $B$ (C04 method) and constant $\zeta_\mathrm{CR}$ NICIL runs, for minimum grain sizes of $0.01\,\mathrm{\mu m}$, $0.1\,\mathrm{\mu m}$ and $1.0\,\mathrm{\mu m}$.
However, the difference was indistinguishable, and thus, the resulting $\eta_\mathrm{AD}$ from NICIL may rely more heavily on the choice of chemical network.
We also checked if the number of grain size bins used affected the results, but still the results did not change.
It is important to note that the derived dust grain properties are that of the disk, and not the envelope. 
Also, there are currently no studies exploring the cosmic-ray ionization rate in the disk or envelope of HOPS-370.
Determining the cosmic-ray ionization rate and dust grain properties in the HOPS-370 protostellar envelope would allow for a better comparison to NICIL. 
Our comparison with NICIL simply represents the closest theoretical scenario we can achieve by using the values derived from observations. 
Therefore, further constraints on the properties of the disk and envelope environment, as well as, comparisons with actual non-ideal MHD simulations should be carried out.

\begin{figure*}[ht!]
  \centering
  \includegraphics[trim={0 0 0 0},clip,width=\textwidth]{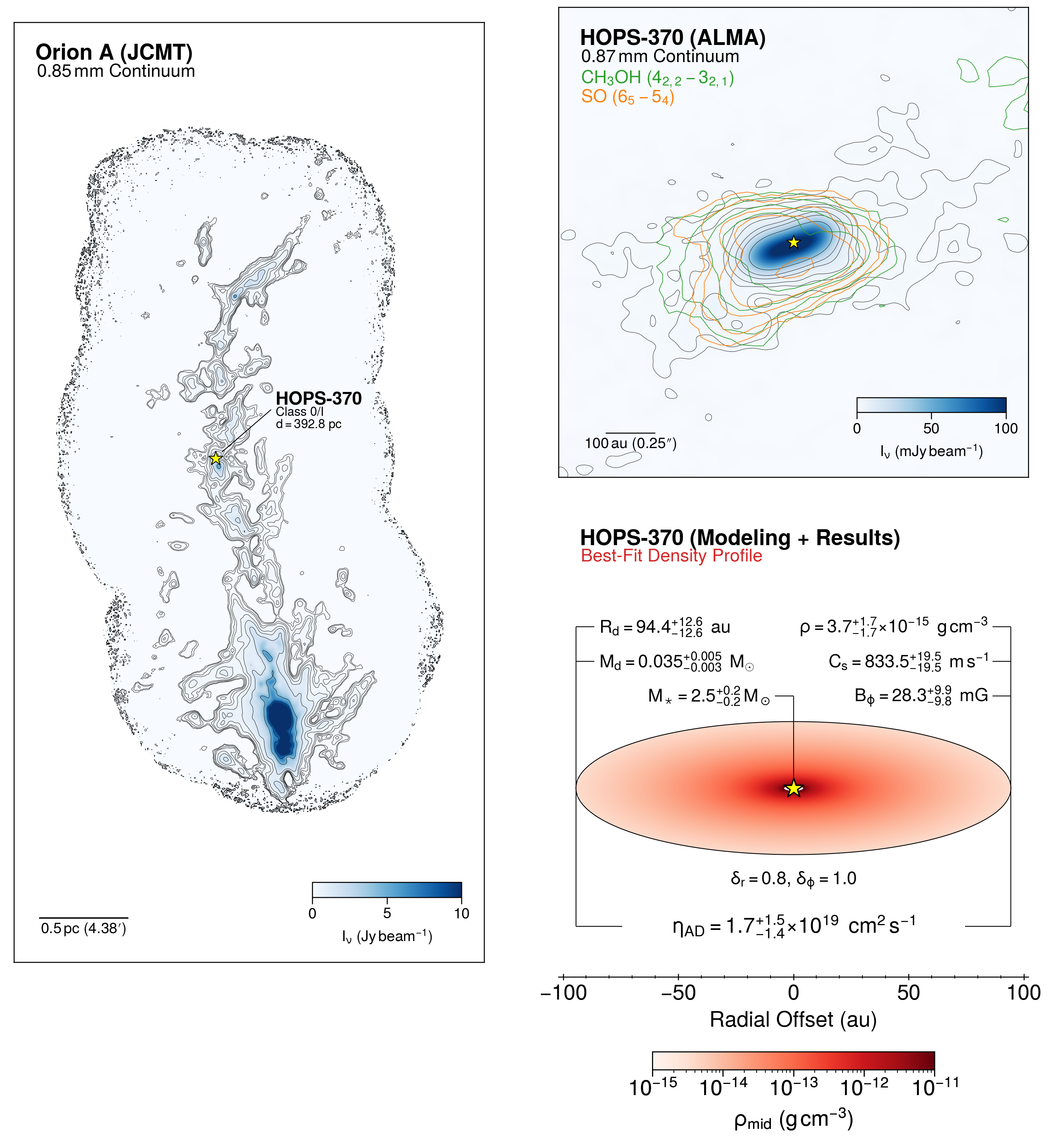}
  \caption{Schematic of the HOPS-370 protostellar system. (Left) 0.85$\,$mm continuum emission of the Orion A molecular cloud taken by the JCMT (Kao \& Yen et al., in prep.).  
  The contour levels shown are 3, 5, 10, 15, 30, 50, 100, 300 and 500$\sigma$, where $\sigma_\mathrm{1.3mm}=15.2\,\mathrm{mJy\,beam^{-1}}$. 
  The location of HOPS-370 is shown with a yellow star, with the protostellar class and distance listed.
  (Right Top) 0.87$\,$mm continuum emission of the protostellar disk around HOPS-370 with self-contour levels of 3, 5, 10, 15, 30, 50 and 100$\sigma$, where $\sigma_\mathrm{0.87mm}=0.39\,\mathrm{mJy\,beam^{-1}}$ \citep{Tobin2020ApJ...905..162T}.
  CH$_3$OH and SO integrated-intensity contours are shown in green and orange, respectively, with contour levels of 3, 5, 10, 15, 30$\sigma$, where $\sigma_\mathrm{CH_3OH}=26.2\,\mathrm{mJy\,beam^{-1}\,km\,s^{-1}}$ and $\sigma_\mathrm{SO}=32.4\,\mathrm{mJy\,beam^{-1}\,km\,s^{-1}}$. 
  These two molecular lines were shown to trace the largest disk radius when modeled together. 
  The position of the continuum peak is marked with a yellow star.
  (Right Bottom) The modeling and results of our ambipolar diffusivity coefficient estimation. 
  The best-fit disk density profile (for $z=0$; i.e. the midplane) is shown in log scale, along with the best-fit quantities used in Equation \ref{eq:2} and our estimated ambipolar diffusivity coefficient at the edge of the disk. 
  }
  \label{fig:schematic}
\end{figure*}

\vspace{-0.1cm}
\section{Conclusion} \label{sec:conclusion}

We present the first estimation of the ambipolar diffusivity coefficient using an analytical equation describing the protostar and disk properties due to ambipolar diffusion. 
We show an illustrative schematic of the HOPS-370 protostellar system to bring together and summarize our results in the context of the multi-scale observations needed for this study (Figure \ref{fig:schematic}). 
The main results of this paper are as follows:
\begin{enumerate}
    \item We derive a generalized analytical expression for the ambipolar diffusivity coefficient in terms of observable quantities in protostellar environments. 
    We show that this relation should be valid, regardless of the global magnetic field orientation with respect to the disk rotation axis.
    \item We make the first estimation of the ambipolar diffusivity coefficient to be $\eta_\mathrm{AD} = 1.7^{+1.5}_{-1.4}\times10^{19}\,\mathrm{cm^{2}\,s^{-1}}$ at the edge of the HOPS-370 protostellar disk, under the assumption that the magnetic field scales with density \citep{Crutcher2004ApJ...600..279C}.
    We use the Alfv\'{e}n speed and Keplerian rotation frequency to estimate the dimensionless Els\"{a}sser number for ambipolar diffusion to be $\mathrm{AM = 1.7^{+1.0}_{-1.0}}$, indicating that ambipolar diffusion is more dynamically important in the region at the edge of the protostellar disk. 
    Estimates of the ambipolar diffusivity coefficient using the inner envelope density, rather than the disk-edge density yields indistinguishable results. 
    \item We use the Non-Ideal MHD Coefficient and Ionisation Library (NICIL) to calculate the non-ideal MHD coefficients using the the physical conditions observed in HOPS-370. 
    We show that the ambipolar diffusivity coefficient from NICIL using various magnetic field strength and cosmic-ray ionization properties is consistent with our derived value. 
    We vary the less certain parameters of $\delta_r$, $\delta_\phi$ and $R_d / R_\mathrm{AD}$ in the ambipolar diffusivity coefficient equation to find the derived value becomes more consistent for decreasing $\delta_r$ and $\delta_\phi$ and increasing $R_d / R_\mathrm{AD}$.
    \item We plot the Ohmic dissipation and Hall effect coefficients along side the ambipolar diffusivity coefficient calculated by NICIL. 
    We find that our derived value shows HOPS-370 lies in an ambipolar diffusion dominated region. 
    This supports the main assumption in the derivation of Equation \ref{eq:1} that ambipolar diffusion is the main diffusion process. 
    When assessing the other assumption made for our derivation of the ambipolar diffusivity equation, we show that they should be valid for HOPS-370. 
    \item We have demonstrated a new methodology for understanding the role of ambipolar diffusion during protostellar disk evolution. Future studies including more sources and more detailed modeling will help to fully understand the role of non-ideal MHD effects in observations of the earliest stages of protostellar disk formation and evolution.
\end{enumerate}

\section*{Acknowledgments}
We thank the anonymous referee for their helpful comments and suggestions on this manuscript. 
This work used high-performance computing facilities operated by the
Center for Informatics and Computation in Astronomy (CICA) at National Tsing Hua University. This equipment was funded by the Ministry of Education of Taiwan, the National Science and Technology Council of Taiwan, and National Tsing Hua University.
S.-P.L. and T.J.T. acknowledge grants from the National Science and Technology Council (NSTC) of Taiwan 106-2119-M-007-021-MY3, 109-2112-M-007-010-MY3 and 112-2112-M-007-011. 
Y.-N.L. acknowledges support from the National Science and Technology Council, Taiwan (NSTC 112-2636-M-003-001) and the grant for Yushan Young Scholar from the Ministry of Education.
S.-J.L. acknowledges the grants from the National Science and Technology Council (NSTC) of Taiwan 111-2124-M-001-005 and 112-2124-M-001-014.
H.-W.Y. acknowledges support from the NSTC grant 110-2628-M-001-003-MY3 and from the Academia Sinica Career Development Award (AS-CDA-111-M03).

\vspace{0.05cm}
\software{Astropy \citep[][\url{http://astropy.org}]{astropy1, astropy2}, asymmetric uncertainty \citep[][]{Gobat2022ascl.soft08005G}, Matplotlib \citep[][\url{http://matplotlib.org/}]{matplotlib}, proplot \citep[][]{luke_l_b_davis_2021_5602155}, Numpy \citep[][\url{http://numpy.org/}]{numpy}}

\appendix
\restartappendixnumbering
\counterwithin{figure}{section}
\counterwithin{table}{section}
\counterwithin{equation}{section}

\twocolumngrid
\section{Derivation of the Ambipolar Diffusivity Coefficient Relation}\label{sec:etaderivation}

The equation describing the disk radius due to ambipolar diffusion first presented by \citet{Hennebelle2016ApJ...830L...8H}, and later by \citet{Lee2021ApJ...922...36L, Lee2024ApJ...961L..28L}, make a number of simplifications. 
Here, we derive a new relationship between the physical properties at the disk-envelope interface to the ambipolar diffusivity coefficient, in order to better compare to more generalized models that are used to fit observations, as in our case for HOPS-370.
We follow the prescriptions given by \citet{Lee2021ApJ...922...36L, Lee2024ApJ...961L..28L}, where more detailed explanations can be found.
We assume ambipolar diffusion is the main diffusion process, as discussed in Section \ref{sec:derivationassumptions}.

First of all, the accretion of angular momentum onto the protostellar disk is counteracted by magnetic braking to rapidly suppress the growth of the disk. 
This results in an equilibrium condition at the disk-envelope interface between the advection and magnetic braking timescales ($\tau_\mathrm{adv} \simeq \tau_\mathrm{br}$) given by 
\begin{gather} 
\tau_\mathrm{adv} \simeq \frac{R}{u_r}, \label{eq:A1}\\
\tau_\mathrm{br} \simeq \frac{\rho u_\phi h}{B_z B_\phi},\label{eq:A2}
\end{gather} 
where $u_r$ and $u_\phi$ are the infalling and rotational velocities, $B_z$ and $B_\phi$ are the poloidal (vertical) and toroidal (azimuthal) magnetic field components, $R$ is the disk radius, $\rho$ is the density at the disk-envelope interface and $h$ is the scale height of the disk at the edge.

Next, $B_\phi$ is generated by the induction of $B_z$ through the differential rotation of the protostellar disk is vertically diffused by ambipolar diffusion.
This results in another equilibrium condition between the generation of $B_\phi$, which happens on the timescale of Faraday induction, and the vertical ambipolar diffusion timescales ($\tau_\mathrm{far} \simeq \tau_\mathrm{diff}$) given by
\begin{gather} 
\tau_\mathrm{far} \simeq \frac{B_\phi h}{B_z u_\phi}, \label{eq:A3}\\
\tau_\mathrm{diff} \simeq \frac{h^2}{\eta_\mathrm{AD}}, \label{eq:A4}
\end{gather} 
where $\eta_\mathrm{AD}$ is the ambipolar diffusivity coefficient.

Since $B_\phi$ should be the dominant component at the protostellar disk edge, we solve our first equilibrium condition for $B_z$ in order to substitute it into our second equilibrium equation, giving 
\begin{equation}
B_z \simeq \frac{\rho u_\phi u_r h}{R B_\phi}. \label{eq:A5}
\end{equation}
Solving the second equilibrium equation in terms of our ambipolar diffusivity coefficient and substituting in our new relation for $B_z$ gives 
\begin{equation}
\eta_\mathrm{AD} \simeq \frac{B_z u_\phi h^2}{B_\phi h} \simeq \frac{\rho u_\phi^2 u_r h^2}{R B_\phi^2}.
\end{equation}
We assume the rotational velocity ($u_r$) and infall velocity ($u_\phi$) both scale with the Keplerian velocity ($v_\mathrm{kep}$) as 
\begin{gather} 
u_r = \delta_r v_\mathrm{kep}, \label{eq:A7}\\
u_\phi = \delta_\phi v_\mathrm{kep}, \label{eq:A8}
\end{gather} 
where $\delta_r$ and $\delta_\phi$ are the the scaling factors and $v_\mathrm{kep}$ defined at the disk edge is
\begin{equation}\label{eq:A9}
v_\mathrm{kep} = \left(\frac{GM}{R}\right)^{1/2}, 
\end{equation}
where $G$ is the gravitational constant and $M = M_\star+M_d$ is the mass of the star+disk system.
From recent MHD simulations of protostellar disk formation including ambipolar diffusion, $u_\phi$ is found to be very close to Keplerian at the disk edge ($\delta_\phi \gtrsim 0.9$), while $u_r$ can be significantly less ($\delta_r \lesssim 0.5$) than the Keplerian velocity, possibly by even a factor of a few \citep[][Section \ref{sec:scalingfactors}]{Lee2021A&A...648A.101L}.
Substituting $u_r$ and $u_\phi$ into our ambipolar diffusivity coefficient relation gives
\begin{equation}
\eta_\mathrm{AD} \simeq \frac{\delta_r \delta_\phi^2 G^{3/2} M^{3/2} \rho h^2}{R^{5/2} B_\phi^2}.
\end{equation}
Assuming vertical hydrostatic equilibrium, the scale height is related to the isothermal sound speed ($C_s$) as
\begin{equation}
h = C_s\left(\frac{R^3}{GM}\right)^{1/2}. \label{eq:A11}
\end{equation}
Now, we can replace $h$ in our ambipolar diffusivity coefficient equation to get a final relation of 
\begin{equation}
\eta_\mathrm{AD} \simeq \frac{\delta_r \delta_\phi^2 G^{1/2} C_s^2 R^{1/2}M^{1/2} \rho}{B_\phi^2}. \label{eq:A12}
\end{equation}
This expression should be valid regardless of the global magnetic field orientation (see Appendix \ref{sec:inclinationeffects} for further discussion).
We have left in the density and sound speed terms, which deviates from the further simplifications made by \citet{Hennebelle2016ApJ...830L...8H} and \citet{Lee2021ApJ...922...36L, Lee2024ApJ...961L..28L}, since these quantities can be modeled for protostellar disks from molecular line observations. 

\section{The Effects of Magnetic Field Inclination on the Ambipolar Diffusivity Coefficient Estimation}\label{sec:inclinationeffects}

For an inclined magnetic field, the equilibrium condition that needs to be satisfied follows as
\begin{equation}
\frac{B_z \sqrt{R}}{\sqrt{\eta_\mathrm{AD} \rho u_r}} + \frac{B_r h}{\sqrt{\eta_\mathrm{AD} \rho u_r R}} = 1,
\end{equation}
where $B_r$ is the radial component of the magnetic field strength and the other symbols have the same meaning as in Appendix \ref{sec:etaderivation} \citep[][]{Lee2024ApJ...961L..28L}.
The relationships between the $B_r$ and $B_z$ components are given by 
\begin{gather} 
B_r = B_0 \left( 2/\pi \right) \sin i \label{eq:B2}\\
B_z = B_0 \cos i \label{eq:B3}
\end{gather} 
where $i$ is the magnetic field inclination with respect to the disk rotation axis ($i=0^\circ$ means the magnetic field direction is aligned/parallel withe disk rotation axis, i.e. the vertical case) and $B_0$ characterizes the amount of magnetic flux that threads the disk region, while the local field strength can be significantly enhanced by magnetic induction due to vertical differential rotation \citep[][]{Lee2024ApJ...961L..28L}. 
Using the relations of $B_z$ and $B_r$, along with Equations \ref{eq:A7}, \ref{eq:A8} and \ref{eq:A11}, we can re-write the previous equation as 
\begin{equation}
\frac{r^{3/4} B_0 \cos i}{\eta_\mathrm{AD}^{1/2} \rho^{1/2} \delta_r^{1/2} G^{1/4} M^{1/4}} + \frac{r^{5/4} B_0 \left(2/\pi\right)\sin i}{\eta_\mathrm{AD}^{1/2} \rho^{1/2} \delta_r^{1/2} G^{3/4} M^{3/4} C_s^{-1}}= 1,
\end{equation}
where $M = M_\star + M_d$ is the total mass of the star+disk system.
Re-writing in terms of $B_0$, we find
\begin{equation}
B_0 = \eta_\mathrm{AD}^{1/2} \rho^{1/2} \delta_r^{1/2} \left[\frac{r^{3/4} \cos i}{ G^{1/4} M^{1/4}} + \frac{r^{5/4} \left(2/\pi\right)\sin i}{ G^{3/4} M^{3/4} C_s^{-1}}\right]^{-1}, \label{eq:B5}
\end{equation}

\quad When the magnetic field is inclined, then  the magnetic field strength derived in Section \ref{sec:bnrelation} should not be assumed as one of the magnetic field components, but rather regarded as the total magnetic field strength. 
We can thus derive an ambipolar diffusivity equation in terms of the total magnetic field strength.
The total magnetic field strength ($B_\mathrm{tot}$) is the sum of squares of all the components written as
\begin{equation}
B_\mathrm{tot}^2 = B_r^2 + B_z^2 + B_\phi^2,
\end{equation}
where $B_r$ and $B_z$ can be substituted again using Equations \ref{eq:B2} and \ref{eq:B3} to give  
\begin{equation}
B_\mathrm{tot}^2 = B_0^2 \left[\left( 2/\pi \sin i \right)^2 + \left( \cos i \right)^2\right] + B_\phi^2. \label{eq:B7}
\end{equation}
Now, we substitute $B_\phi$ using our derived relationship in Equation \ref{eq:A12} to get
\begin{equation}
\begin{split}
B_\mathrm{tot}^2 &= B_0^2 \left[\left( 2/\pi \sin i \right)^2 + \left( \cos i \right)^2\right] \\
&+ \delta_r \delta_\phi^{2} G^{1/2} C_s^{2} R^{1/2} M^{1/2} \rho \eta_\mathrm{AD}^{-1}. \label{eq:B8}
\end{split}
\end{equation}

We now have two equations (\ref{eq:B5} and \ref{eq:B8}) with two unknowns ($B_0$ and $\eta_\mathrm{AD}$). 
Thus, we substitute Equation \ref{eq:B5} into Equation \ref{eq:B8} to remove $B_0$ and obtain a second-order polynomial ambipolar diffusivity equation for an inclined magnetic field of
\begin{equation}
\begin{split}
&\eta_\mathrm{AD}^{2} \rho \delta_r\left[\frac{r^{3/4} \cos i}{ G^{1/4} M^{1/4}} + \frac{r^{5/4} \left(2/\pi\right)\sin i}{ G^{3/4} M^{3/4} C_s^{-1}}\right]^{-2} \\ &\times \left[\left( 2/\pi \sin i \right)^2 + \left( \cos i \right)^2\right] \\
&- \eta_\mathrm{AD} B_\mathrm{tot}^2 \\
&+ \delta_r \delta_\phi^{2} G^{1/2} C_s^{2} R^{1/2} M^{1/2} \rho = 0,
\end{split}
\end{equation}
which can be solved to find the ambipolar diffusivity coefficient.
Using the same values as in Section \ref{sec:firstestimation} and a magnetic field inclination with respect to the disk rotation axis of $45\pm22^{\circ}$ \citep{Yen2021ApJ...916...97Y}, we find $\eta_\mathrm{AD} = 1.798^{+0.042}_{-0.007}\times10^{19}\,\mathrm{cm^{2}\,s^{-1}}$, where the reported errors are only due to the error on the magnetic field inclination angle. 
For one, this value is extremely close to and within error of the value previously derived in Section \ref{sec:firstestimation}.
Additionally, the errors due to only the magnetic field inclination are a few orders of magnitude smaller than in the previously derived value.
This show that the magnetic field inclination has essentially no effect on our derived ambipolar diffusivity coefficient.

For completeness, we check the corresponding values of $B_0$, $B_z$, $B_r$ and $B_\phi$ to see if which component of the magnetic field is dominant. 
We use Equations \ref{eq:B2}, \ref{eq:B3}, \ref{eq:B5} and \ref{eq:B7} to estimate values of $B_0 \approx 5.5\,\mathrm{mG}$, $B_z \approx 3.9\,\mathrm{mG} $, $B_r \approx 2.5\,\mathrm{mG}$ and $B_\phi \approx 28.0\,\mathrm{mG}$.
This shows that the $B_\phi$ component of the magnetic field still dominates even when considering the orientation.
Thus, our derived relation is considered to be generalized and it is correct to assume $B_\phi \approx B_\mathrm{tot}$ in our initial assumptions (Section \ref{sec:bnrelation}). 

\section{Ambipolar Diffusivity Coefficient Estimation using Inner Envelope Density}\label{sec:etaADenvelope}

\begin{figure}[t!]
  \includegraphics[width=\columnwidth]{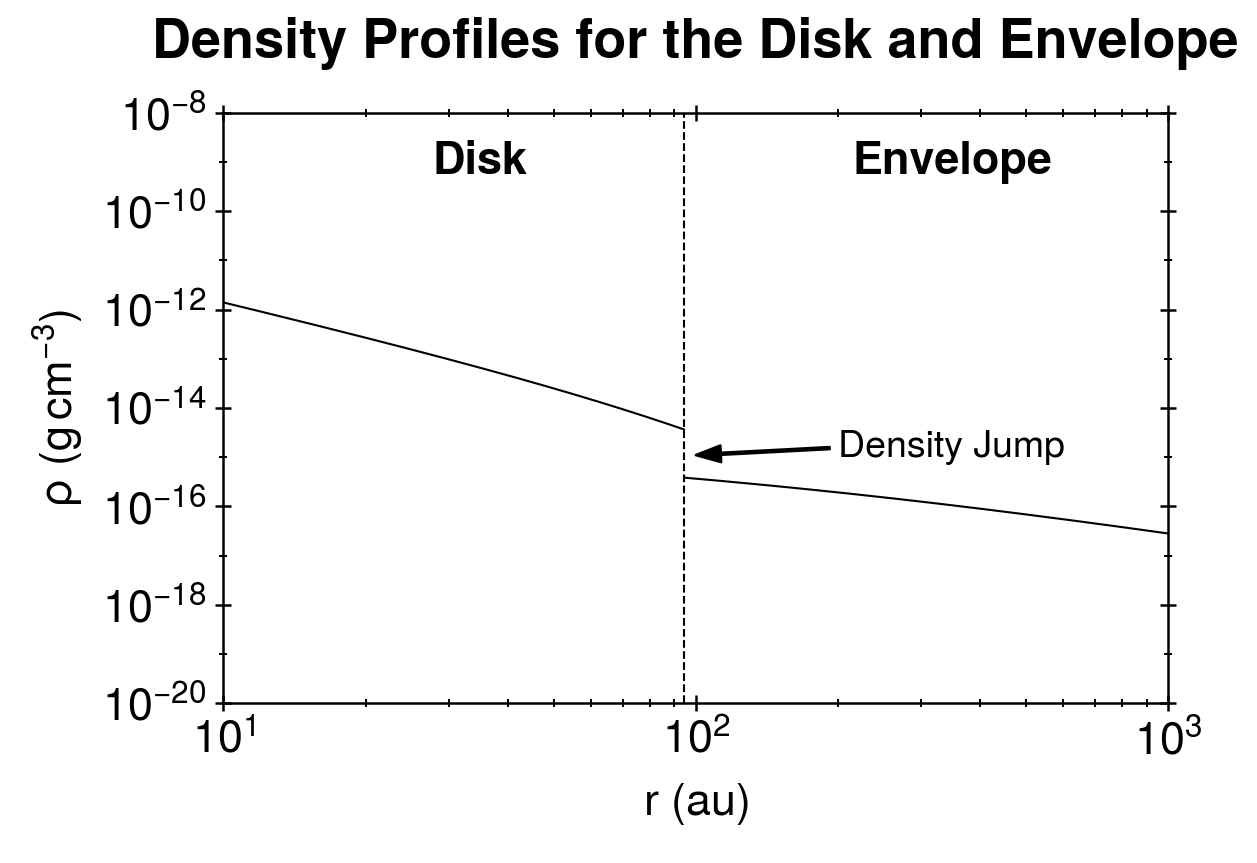}
  \caption{Comparison between the best-fit disk and envelope volume density profiles for HOPS-370 \citep{Tobin2020ApJ...905..162T}. The vertical dashed line represents the radius of the Keplerian gas disk.}
  \label{fig:etaADenvelopecompare}
\end{figure}

As described in Section \ref{sec:density}, we could equally derive a density at the disk-envelope interface from the best fit envelope volume density relation in \citet{Tobin2020ApJ...905..162T}. 
We stress that much of the envelope emission may be resolved-out, which could effect the fitting results.
However, it is interesting to still investigate how the ambipolar diffusivity coefficient estimation is effected using the current best-fit results.
To model the envelope emission using the molecular line data, \citet{Tobin2020ApJ...905..162T} uses the following relation for the envelope
\begin{equation}
\begin{split}
    \rho_\mathrm{env} (r) &= \frac{\dot{M}_\mathrm{env}}{4 \pi} \left(G M_\star r^3\right)^{-1/2} \\
    &\times\left(1+\frac{\mu}{\mu_0}\right)^{-1/2} \left(\frac{\mu}{\mu_0}+2\mu_0^2\frac{R_c}{r}\right)^{-1},
\end{split}
\end{equation}
where $\dot{M}_\mathrm{env}$ is the envelope-to-disk mass-accretion rate, $\mu_0 = \cos\theta_0$ is the initial polar angle of a streamline trajectory out to $r \rightarrow \infty$, $\mu = \cos\theta$ is the polar angle along the streamline trajectory, $R_c$ is the centrifugal radius where the infalling material has sufficient angular momentum to maintain an orbit the central protostar.
We take the simplified case in the mid-plane of the inner envelope, where $\theta_0=\theta=90^\circ$, which simplifies the equation to 
\begin{equation}\label{eq:C2}
    \rho_\mathrm{env} (r) = \frac{\dot{M}_\mathrm{env}}{4 \pi} \left(2G M_\star r^3\right)^{-1/2} \left(1+2\frac{R_c}{r}\right)^{-1}.
\end{equation}
We show the best-fit disk and envelope density profiles from \citet{Tobin2020ApJ...905..162T} in Figure \ref{fig:etaADenvelopecompare}.
We see a clear difference in densities between the disk and envelope (density jump).
We plug in $\dot{M}_\mathrm{env} = 3.2\pm0.6\times10^{-5}\,M_\odot\,\mathrm{yr^{-1}}$ (no error bars are reported, so we assume a 20\% error), $M_\star = 2.5\pm0.2\,M_\odot$, $r = R_c = R_d = 94.4\pm12.6\,\mathrm{au}$ \citep[][]{Tobin2020ApJ...905..162T}, we find $\rho_\mathrm{env} (r) = 3.8\pm1.2\times10^{-16}\,\mathrm{g\,cm^{-3}}$.
Using our new inner envelope density, we re-apply the same steps as in Section \ref{sec:bnrelation} to scale the magnetic field from the core-scale density using the C04 method. 
This gives us newly estimated magnetic field strength of $B_{\mathrm{tot},e} = 9.2\pm2.8\,\mathrm{mG}$.
We now plug in the values of
\begin{gather*} 
    \delta_r = 0.8, \\
    \delta_\phi = 1.0, \\
    C_s = 833.0^{+19.5}_{-19.5}\,\mathrm{m\,s^{-1}}, \\
    R_d = 94.4^{+12.6}_{-12.6}\,\mathrm{au}, \\
    M_\star = 2.5^{+0.2}_{-0.2}\,M_\odot, \\
    M_d = 0.035^{+0.005}_{-0.003}\,M_\odot \\
    \rho_d=3.8^{+1.2}_{-1.2}\times10^{-16}\,\mathrm{g\,cm^{-3}} \\
    B_\phi = 9.2^{+2.8}_{-2.8}\,\mathrm{mG}, 
\end{gather*}
into the ambipolar diffusivity coefficient equation (Equation \ref{eq:3}) to obtain 
\begin{align*}
    \eta_\mathrm{AD} &= 1.7^{+1.2}_{-1.2}\times10^{19}\,\mathrm{cm^{2}\,s^{-1}}\\
    &= 2.4^{+1.6}_{-1.6}\times10^{-1}\,\mathrm{s}.
\end{align*}
The dimensionless Els\"{a}sser number is estimated to be $\mathrm{AM} = 1.7^{+0.8}_{-0.8}$ (C04 method).
These results are indistinguishable from the values calculated using the disk edge quantities. 
Even though the envelope density is estimated to be an order of magnitude lower than the disk edge, the magnetic field strength is also lower as a result.
Since $\eta_\mathrm{AD}$ has a dependence on $\sim \rho$ and $\sim B_\phi^{-2}$, the values end up offsetting each other to give similar estimates.
This shows that either the disk or envelope density can be used interchangeably to obtain a value for the ambipolar diffusivity coefficient. 
We again check whether $B_\phi > B_z$ using Equation \ref{eq:A5}, and find $B_z \approx 1.38\,\mathrm{mG}$.
Thus, $B_\phi$ is still the dominant component, although it is more comparable to $B_z$ in this case.

\vspace{0.5cm}
\section{Magnetic Field Strength Estimation}\label{sec:etaADL24}
Recently, \citet{Lee2024ApJ...961L..28L} derive a new analytical expression to describe how the magnetic field should scale with density inside a collapsing protostellar envelope for the first time. 
Considering the two density regimes ad the core and disk scales, their magnetic field - density relation (Equation C7 in their paper) can be simplified to
\begin{equation}\label{eq:13}
    B_\mathrm{0,d} = B_\mathrm{0,c} \left(\frac{M_\star+M_\mathrm{d}}{M_\mathrm{c}}\right)^{0.25}\left(\frac{\rho_\mathrm{d}}{\rho_\mathrm{c}}\right)^{0.525},
\end{equation}
which can also be used to scale the magnetic field strength down to inner envelope/protostellar disk density regimes (hereafter, referred to as the L24 method).
Here $B_0$ has the same meaning as in Appendix \ref{sec:inclinationeffects} and characterizes the magnetic flux threading the disk. 
We assume the case of a vertical magnetic field, since the effects of inclination are minimal on our estimates, which gives $B_0 \approx B_z$.
Plugging in our known values of $B_\mathrm{tot,c}$, $M_\star$, $M_\mathrm{d}$, $\rho_\mathrm{d}$ and $\rho_\mathrm{c}$, we estimate $B_\mathrm{0,d} \approx B_z = 17.6^{+6.3}_{-6.2}\,\mathrm{mG}$.
This is compatible with the calculations in the main text within order of magnitude. 
The discrepancy results from model presumptions that require further examination, while we do not discuss the details.

\bibliography{references}
\bibliographystyle{aasjournal}

\end{document}